\documentclass[]{aa}  

\usepackage[utf8]{inputenc}
\usepackage{graphicx}
\usepackage{txfonts}
\usepackage{hyperref}
\usepackage{lscape}

\newcommand{\msun}{M_\odot}

\begin{document} 

    \title{
    Type Ic supernovae from the (intermediate) Palomar Transient Factory}
    \titlerunning{SNe Ic from (i)PTF}

      \author{C.~Barbarino \inst{1} \href{https://orcid.org/0000-0002-3821-6144}{\includegraphics[scale=0.5]{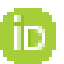}}
      \and J.~Sollerman \inst{1} \href{https://orcid.org/0000-0003-1546-6615}{\includegraphics[scale=0.5]{ORCIDiD_icon16x16.eps}}
        \and F.~Taddia \inst{1}
        \and C.~Fremling \inst{2}
        \and E.~Karamehmetoglu\inst{3}
        \and I.~Arcavi\inst{4,5} \href{https://orcid.org/0000-0001-7090-4898}{\includegraphics[scale=0.5]{ORCIDiD_icon16x16.eps}}
        \and A.~Gal-Yam\inst{6} \href{https://orcid.org/0000-0002-3653-5598}{\includegraphics[scale=0.5]{ORCIDiD_icon16x16.eps}}
        \and R. Laher\inst{7}
        \and S.~Schulze\inst{6} \href{https://orcid.org/0000-0001-6797-1889}{\includegraphics[scale=0.5]{ORCIDiD_icon16x16.eps}}
        \and P.~Wozniak\inst{8}
        \and Lin~Yan\inst{9} \href{https://orcid.org/0000-0003-1710-9339}{\includegraphics[scale=0.5]{ORCIDiD_icon16x16.eps}}
    }    
	\institute{The Oskar Klein Centre, Department of Astronomy, Stockholm University, AlbaNova, 10691 Stockholm, Sweden\\ \email{cristina.barbarino@astro.su.se}
	\and Division of Physics, Mathematics, and Astronomy, California Institute of Technology, Pasadena, CA 91125, USA
    \and Department of Physics and Astronomy, Aarhus University, Ny Munkegade 120, DK-8000 Aarhus C, Denmark
    \and The School of Physics and Astronomy, Tel Aviv University, Tel Aviv 69978, Israel
    \and CIFAR Azrieli Global Scholars program, CIFAR, Toronto, Canada
    \and Department of Particle Physics and Astrophysics, Weizmann Institute of Science, Rehovot 76100, Israel
    \and IPAC, California Institute of Technology, 1200 E. California Blvd, Pasadena, CA 91125, USA
    \and Los Alamos National Laboratory, MS-D466, Los Alamos, NM 87545, USA
    \and The Caltech Optical Observatories, California Institute of Technology, Pasadena, CA 91125, USA
   }

   \date{Received; accepted}

 
  \abstract
   {Type Ic supernovae represent the explosions of the most stripped massive stars, but their progenitors and explosion mechanisms remain unclear. Larger samples of observed supernovae can help characterize the population of these transients.}
   {We present an analysis of 44 spectroscopically normal Type Ic supernovae, with focus on the light curves. The photometric data were obtained over 7 years with the Palomar Transient Factory (PTF) and its continuation, the intermediate Palomar Transient Factory (iPTF). This is the first homogeneous and large sample of SNe Ic from an untargeted survey, and we aim to estimate explosion parameters for the sample.}
   {We present K-corrected $Bgriz$ light curves of these SNe, obtained through photometry on template-subtracted images. We performed an analysis on the shape of the $r$-band light curves and confirmed the correlation between the rise parameter $\Delta m_{-10}$ and the decline parameter $\Delta m_{15}$. 
   Peak $r$-band absolute magnitudes have an average of $ -17.71 \: \pm \: 0.85$ mag. To derive the explosion epochs, we fit the $r$-band lightcurves to a template derived from a well-sampled light curve.
   We computed the bolometric light curves using $r$ and $g$ band data, $g-r$ colors and bolometric corrections. Bolometric light curves and \ion{Fe}{II}~$\lambda$5169 velocities at peak were used to fit to the Arnett semianalytic model in order to estimate the ejecta mass $M_{ej}$, the explosion energy $E_{K}$ and the mass of radioactive nickel $M({^{56}}\mathrm{Ni})$ for each SN.}
   {Including 41 SNe, we find average values of $<M_{ej}>=4.50\pm0.79 \; \msun$,  $<E_{K}>=1.79 \pm 0.29 \; 
   ~\mathrm{\times10^{51}}$~erg, and $<M_{^{56}\mathrm{Ni}}>= 0.19 \pm 0.03~\msun$.
   The explosion-parameter distributions are comparable to those available in the literature, but our large sample also includes some transients with narrow and very broad light curves leading to more extreme ejecta masses values.
   }
   {}


\maketitle

\section{Introduction}\label{Introduction}

Core-collapse supernovae (CC SNe) are the explosions of massive stars ($\gtrsim 8 \: \msun$) which undergo gravitational collapse of the core at the end of their lifes. Their classification relies on the presence/absence of some spectroscopic features \citep[e.g.,][]{Filippenko97,GalYam17}.

When they lack, partially or totally, hydrogen (H) or helium (He) they are called stripped envelope SNe (SE SNe). In this class we find SNe IIb, Ib and Ic. Type Ib SNe show no H but He in their spectra. SNe IIb show an initial signature of H at peak, which then disappears over time as their spectra become similar to those of Type Ib SNe. Supernovae Type Ic are the ones which lack both H and He. Two main scenarios have been proposed as progenitor systems of SE SNe: i) single and massive Wolf-Rayet (WR) stars that lose their outer envelopes through radiation-driven stellar winds (\citealt{Begelman86}; \citealt{Woosley95}), and ii) lower mass stars in binary systems characterized by mass transfer (\citealt{Wheeler85}; \citealt{Podsiadlowski92}; \citealt{Yoon10}).
It is still a matter of debate whether one or both of these progenitor channels can explain the observed SE SN population.

The estimated masses of the ejecta for SE SNe seem to favour the lower mass binary star scenario, rather than very massive single WR stars (\citealt{Eldridge13}, \citealt{Lyman16}, \citealt{Cano13}); and for example
the data for the individual Type Ib SN iPTF13bvn seem to be more consistent with a binary system \citep{Cao13,Bersten14,Fremling14,Fremling16, Eldridge15}.

However, some SE SNe do seem to be originating from very massive stars (e.g. SN 2011bm, \citealt{Valenti12}; OGLE-2014-SN-131, \citealt{Emir17}; LSQ14efd, \citealt{Barbarino17}) and thus favour a single progenitor system.

To complicate the picture we can have broad$-$lined Type Ic SNe that shows spectra of SNe Ic (e.g. SN 2003jd, \citealt{Valenti08}) as well as super-luminous SNe that spectroscopically resemble SNe Ic (e.g. SN 2010gx, \citealt{Pastorello10}).

The photometric samples available in the literature either refer to SE SNe (\citealt{Lyman16}; \citealt{Prentice16}; \citealt{Taddia18b}, \citealt{Prentice19}) or to SNe Ibc (\citealt{Drout11}; \citealt{Taddia15}), including both SNe Ib and Ic. A major sample of merely SNe Ic (or Ib) is not available.
Thanks to the untargeted surveys, the Palomar Transient Factory (PTF, \citealt{Rau09}; \citealt{Law09}) and its continuation, the intermediate Palomar Transient Factory (iPTF, \citealt{Kulkarni13}), we can here present 
optical observations of 
a large (44 objects) and homogeneous sample of spectroscopically normal SNe Ic. This enables a study of the properties of the SN population and their progenitor stars. \\
The paper is organized as follow: in Sect.~\ref{Sect:sample} we present the sample while the photometry and the data reduction are presented in Sect.~\ref{Sect:phot&red}. The analysis of the light curves is discussed in Sect.~\ref{Sect:LC}. Bolometric light curves are presented in Sect.~\ref{Sect:bolo}. The spectra and the analysis of the velocities are shown in Sect.~\ref{Sect:spectra}. The explosion parameters are estimated in Sect.~\ref{Sect:ExplPar}. The results are discussed, summarized and compared with the literature in Sect.~\ref{Sect:discussion}.

\section{The sample} \label{Sect:sample}

\begin{figure*}
\centering
 \includegraphics[width=10cm]{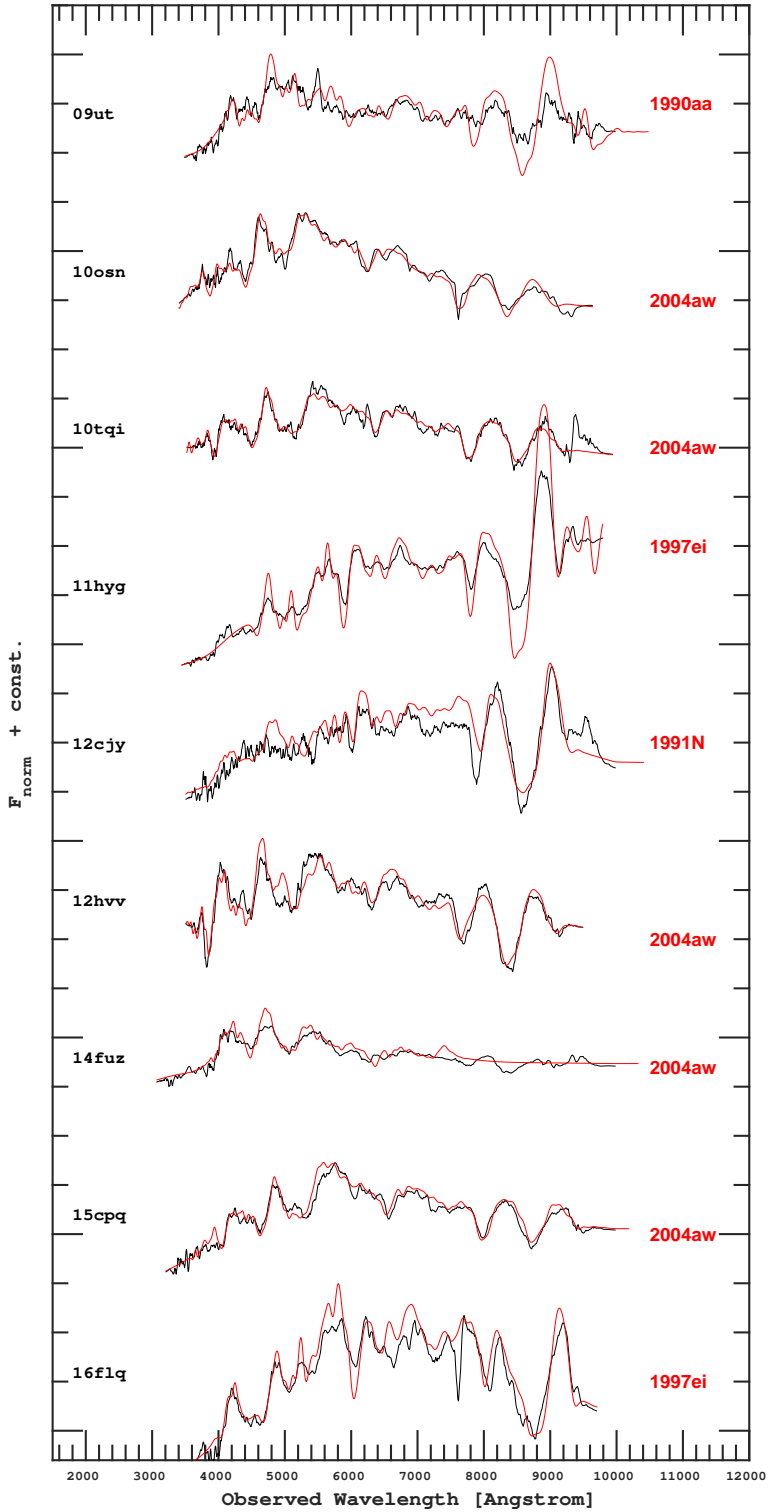}
 \caption{Examples of 9 SNe from the sample and their classification through the SNID package.}
 \label{Fig:SpecClass}
 \end{figure*}

The SN sample presented in this work consists of 60 SNe Ic and 17 SNe Ibc discovered and followed by PTF and iPTF. The classification of these objects was based on spectroscopy and performed using the Supernova Identification code (SNID, \citealt{snid}) with the addition of SE SN templates from \citet{Modjaz14}. For illustration, the classification spectra for nine SNe of the sample are shown in Fig.~\ref{Fig:SpecClass}. The SNe have been selected in order to illustrate the variety of data quality of the spectral sample. We also note that that details provided by SNID, such as phases and redshifts are in agreement with the actual value for each SN within the uncertainties.

\begin{figure}
 	\includegraphics[width=8.5cm]{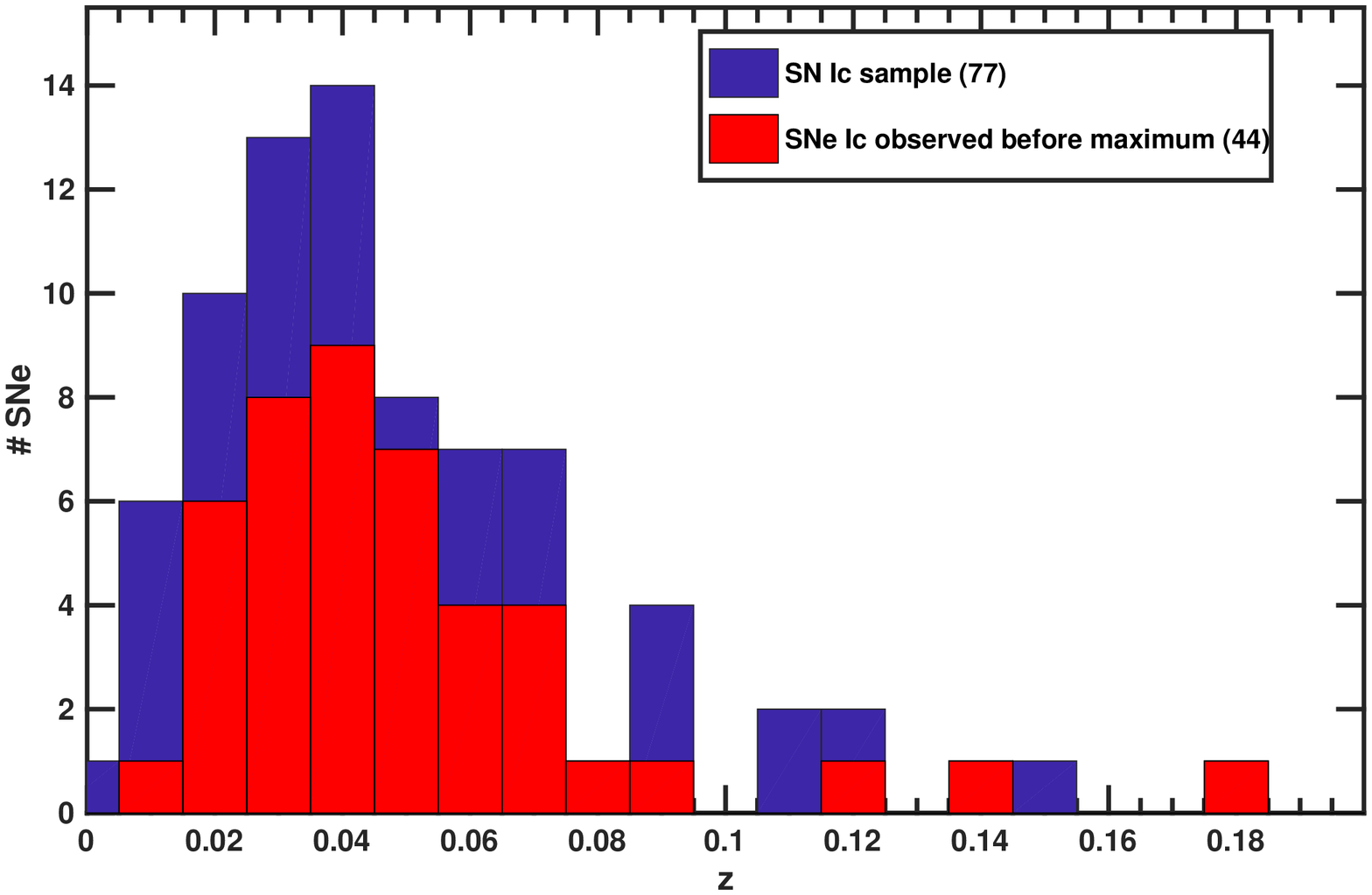}
 	\caption{Redshift distribution of all (i)PTF SNe Ic (blue) and of the 44 SNe Ic discovered before peak (red). The latter constitute the sample analysed in this work.}
 	\label{Fig:RedDist}
 \end{figure}

The objects classified as Type Ibc here are those for which a clear difference between Ib and Ic could not be established, due to the spectral-data quality. This represents the full sample of the PTF + iPTF (hereafter combined into (i)PTF) Type Ic population and our classifications are mostly consistent with those by \citet{Fremling18}.
The redshift of the host galaxy has been adopted when 
available\footnote{The values refer to the ones available at the NASA/IPAC extragalactic database; {http://ned.ipac.caltech.edu}}.
When this information was not available, the redshift was estimated from the host-galaxy lines when detected, 
otherwise we adopted the best fit from SNID.
The redshifts of the sample span the interval $z= 0.004486 - 0.176$. 
The mean value is $z= 0.049 \pm 0.033$.
The redshift distribution is presented in Fig.~\ref{Fig:RedDist}. 

\begin{figure*}
	\centering
		\includegraphics[width=18cm]{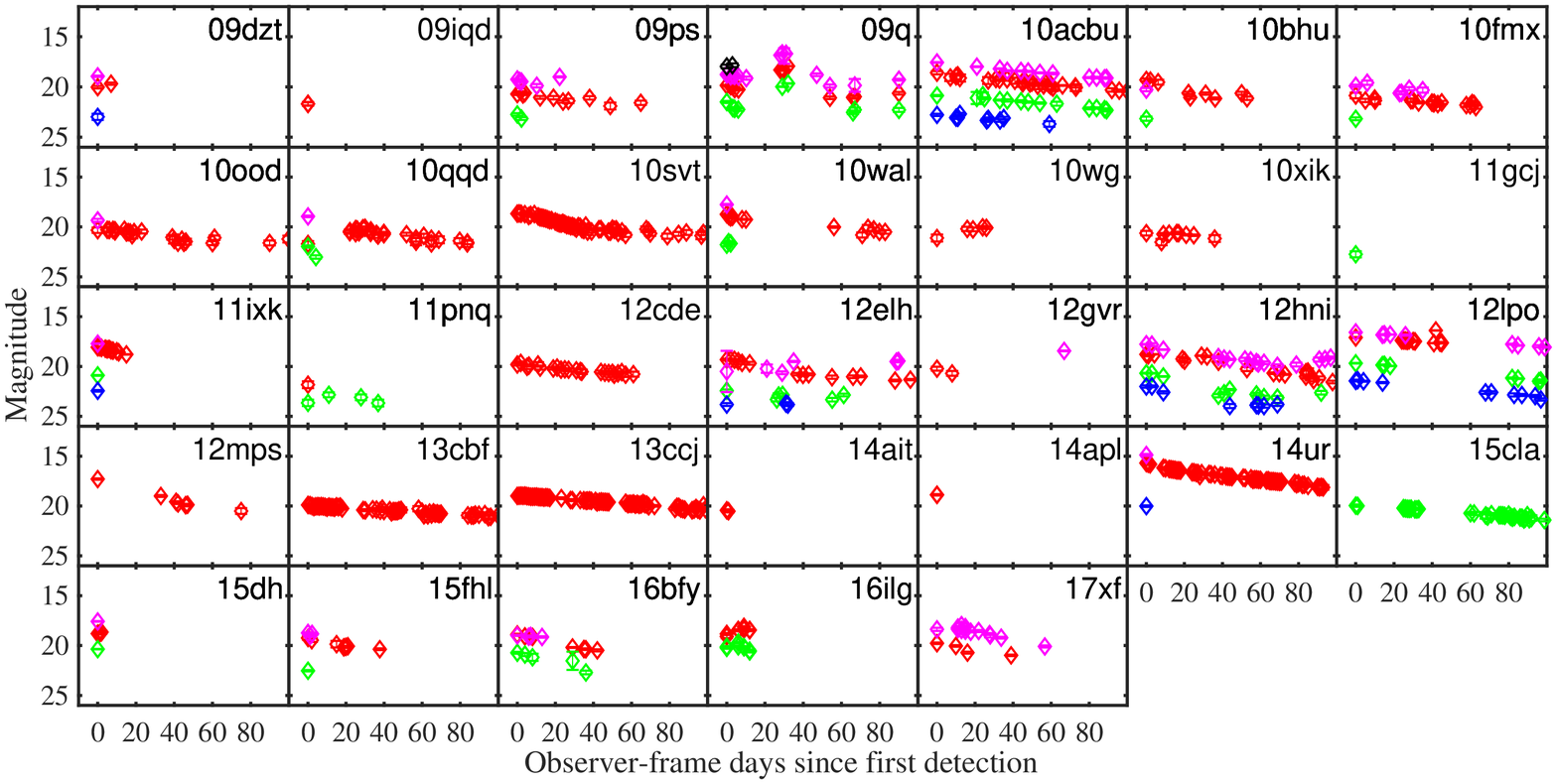}
	\caption{Light curves for all the 33 SNe Ic and Ibc in (i)PTF which do not show an observed peak and will not be included in the overall analysis. We plot the apparent magnitude as a function of time since first detection, in the observer's frame. Shifts have been applied for clarity as indicated in the legend in the bottom row.}
	\label{Fig:LCs}
\end{figure*}

The redshift was used to compute the luminosity distance for each SN. We adopted the WMAP 5-year \citep{Komatsu09} cosmological parameters $H_{0} = 70.5  \:  \mathrm{km \:  s^{-1}} \: \mathrm{Mpc}^{-1}$, $\Omega_{M} = 0.27$, $\Omega_{\Lambda} = 0.73$ and corrections for peculiar motions (Virgo, GA, Shapley) are included using a function by \citet{nyholm20} based on \citet{mould00a,mould00b} and the NASA/IPAC Extragalactic Database velocity routine.. With these assupmtions, the distance of the sample ranges in the interval D = $19.1-847.6 \: \mathrm{Mpc}$. \\
The Milky Way extinction was obtained from \cite{Schlafly11}\footnote{via the NASA/IPAC infrared science archive; {https://irsa.ipac.caltech.edu/applications/DUST/}}.
The treatment of the host extinction is presented in Sect.~\ref{Sect:ColHost}.
Our sample was observed mainly in the $r$ and $g$ bands, with some photometric data also in the $B$, $i$ and $z$ bands for some objects.
All the light curves in apparent magnitudes for the 77 SNe are shown in Figs.~\ref{Fig:LCs} and ~\ref{Fig:LC_Kcorr}. 

Among the 60 SNe Ic + 17 SNe Ibc of our sample, 
44 were observed before peak in at least one band. 
In the following analysis we will focus only on these latter 44 objects.
These SNe were observed in the $r$ band with an average cadence of 3 days and they have been followed with a median coverage of 66 days post peak.
The 44 SNe in the sample have a redshift in the interval $z= 0.01377 - 0.176$, the mean value being $z= 0.051 \pm 0.032 $.
The redshifts for the SNe are 
highlighted in Fig.~\ref{Fig:RedDist}
and
listed in Table~\ref{Table:SNprop} where it is specified if the redshift was obtained with SNID or measured from narrow emission lines. They are all provided with 3 decimals.

This paper is mainly focused on the photometric data of the sample. However, each SN has at least one spectrum obtained by the (i)PTF survey and collaborators. The analysis of these spectra was published in \citet{Fremling18}.
This (i)PTF data set of SNe Ic is thus unique given its untargeted nature, its large size, its early coverage, its high cadence and multiband coverage.
A detailed analysis of the host galaxies of many of these Type Ic SNe, and of the hosts of the PTF sample of broad lined Type Ic SNe (SNe Ic-BL), was presented by \citet{Modjaz20}. We compared the classification reported in \citet{Modjaz20} with the one we present in this work. We notice that PTF09ps and PTF10bip are presented as Ic/Ic-BL, while PTF11gcj is classified as Ic-BL. We consider these three as SNe Ic as our classification suggested.
The analysis of the (i)PTF sample of SNe Ic-BL was published by \citet{Taddia19b}.
Here we focus on the spectroscopically normal (as opposed to broad lined) SNe Ic.
Some of the SNe in our sample have already been studied in other works. SNe PTF09dh, PTF11bli, PTF11jgj, PTF11klg, PTF11rka and PTF12gzk were presented in \citet{Prentice16}. In that work  PTF09dh was classified as a SN Ic-BL, but we include it here since we have re-classified it as a spectroscopically-normal SN Ic. We also note that PTF10vgv, presented as a SN Ic in \citet{Prentice16},  is not included in this work since we re-classified it as a SN Ic-BL; therefore, PTF10vgv was presented within the sample of SNe Ic-BL from (i)PTF in \citet{Taddia19b} and also in \citet{Corsi12}.
PTF12gzk was presented and discussed in \citet{BenAmi12}; PTF11bov is also known as SN~2011bm and was studied by \citet{Valenti12} and \citet{Taddia16}; iPTF12gty was presented in \citet{DeCia17} and also dicussed by \citet{Quimby18} as a superluminous supernova of Type I (SLSN-I). However, our spectroscopic classification suggests some similarity with a SN Ic so we included it in this work. iPTF15dtg was presented and discussed in \citet{Taddia16} and late-time data are shown in \citet{Taddia19a}. PTF11mnb was presented in a separate paper as a SN Ic \citep{Taddia18a} but it is also discussed in \citet{Quimby18} as a possible SLSN-I. We include PTF11mnb in this work since our spectroscopic classification agrees with that of a SN Ic by \citet{Taddia18a}. Finally, iPTF14gqr was presented in \citet{De18}.\\
We notice that some SNe, spectroscopically classified as normal SNe Ic, show quite broad light curves compared to the bulk of the 
sample\footnote{PTF11mnb, PTF11rka, PTF12gty, iPTF15dtg, iPTF16flq and iPTF16hgp.}.
In order to identify these SNe, the SE SN template presented by \citet{Taddia15} was used as a reference. The template was shifted and stretched to fit the SN light curve at maximum. The SNe that presented a stretch factor higher than 1.5 were considered as having a broad light curve. The method used and these SNe will be discussed in detail in a forthcoming paper (Karamehmetoglu et al. in prep).  Our sample also includes an object with a very narrow light curve\footnote{iPTF14gqr.}. The presence of a wider variety of of objets is likley due to large sample and the untargeted nature of the survey.

\section{Photometric observations and data reduction} \label{Sect:phot&red}
 
SN discovery and early photometric observations were performed with the 48-inch Samuel Oschin Telescope at Palomar Observatory (P48), equipped with the 96 Mpx mosaic camera CFH12K \citep{Rahmer08}, a Mould $r$-band filter \citep{Ofek12} and a $g$-band filter.  For 34 
SNe in our sample, further follow-up was performed with the automated Palomar 60-inch telescope (P60, \citealt{Cenko06}), often in $Bgri$ bands.
Point spread function (PSF) photometry was obtained on template subtracted images using the Palomar Transient Factory Image Differencing and Extraction (PTFIDE) pipeline \citep{Masci17} for P48 data and the FPipe pipeline presented in \citet{Fremling16} for the P60 data. The photometry was calibrated using  Sloan Digital Sky Survey (SDSS) stars \citep{Ahn14} in the SN field. All light curves will be released.
 
 \section{Supernova light curves} \label{Sect:LC}
 
 \begin{figure*}
 	\centering
 	\includegraphics[width=18cm]{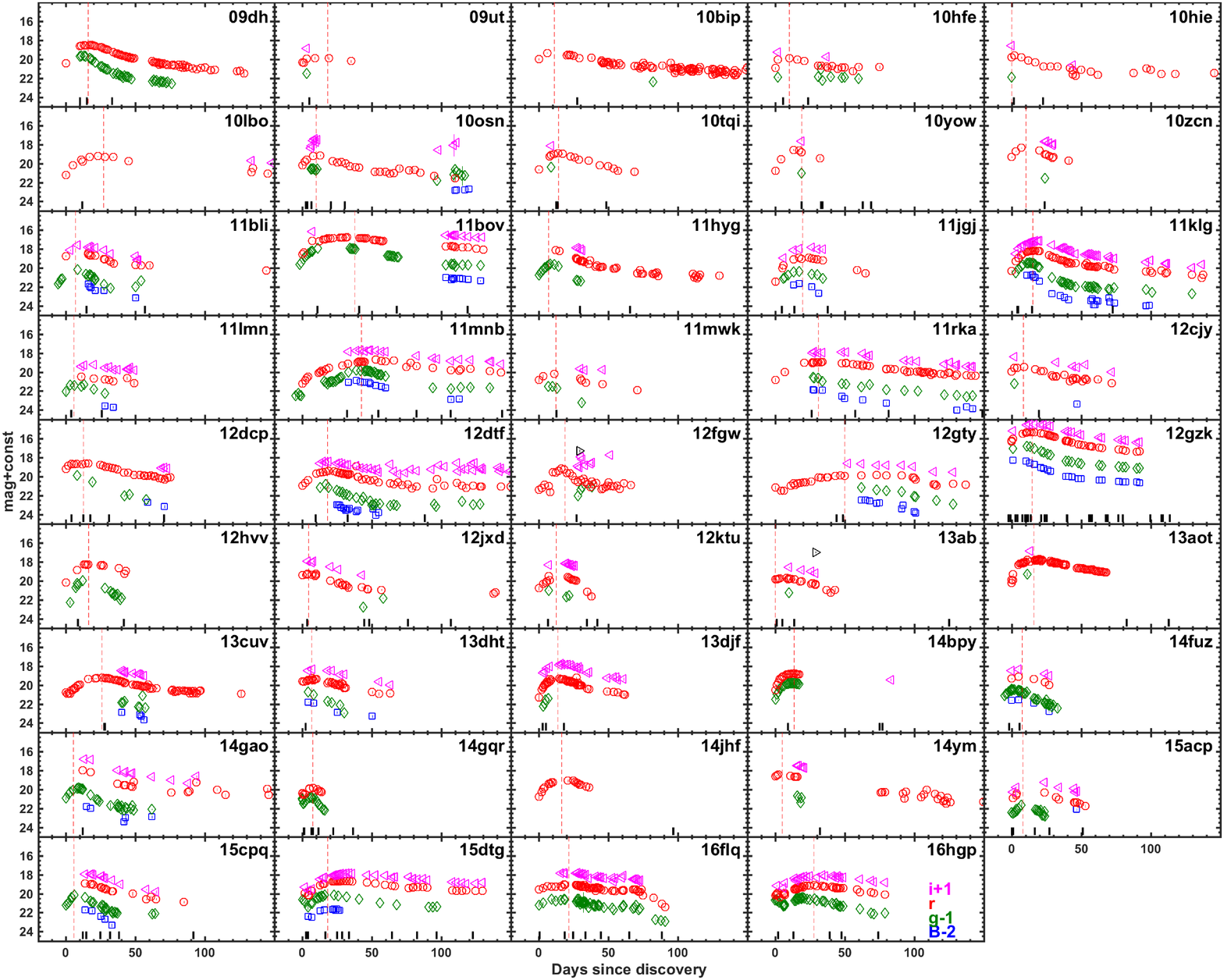}
 	\caption{Light curves in $B,g,r,i$ of the 44 SNe Ic for which we have  pre-maximum observations. We plot the apparent magnitude as a function of days since discovery.  Shifts have been applied for clarity, as indicated in the legend in the bottom row. The peak epoch is shown as a dashed red line. The black dashed lines at the bottom represent epochs of spectral observations.}
 	\label{Fig:LC_Kcorr}
 \end{figure*}
 
In Fig.~\ref{Fig:LC_Kcorr} we present the photometric observations in the optical bands available for our SN sample. As mentioned before, we will proceed with our analysis only for those 44 SNe which have been observed before peak. The observed light curves were first corrected for time dilation and K-corrections.

 \subsection{K-corrections}
 
Most of the SNe were observed in the $r$ band and we 
estimated the observed peak epoch in the $r$ band ($t_{r}^{max}$) by fitting a polynomial to the $r$-band light curves. The peak epoch is shown with a dashed red line in Fig.~\ref{Fig:LC_Kcorr}.
 
\begin{figure*}
 	\centering
 	\includegraphics[width=10cm]{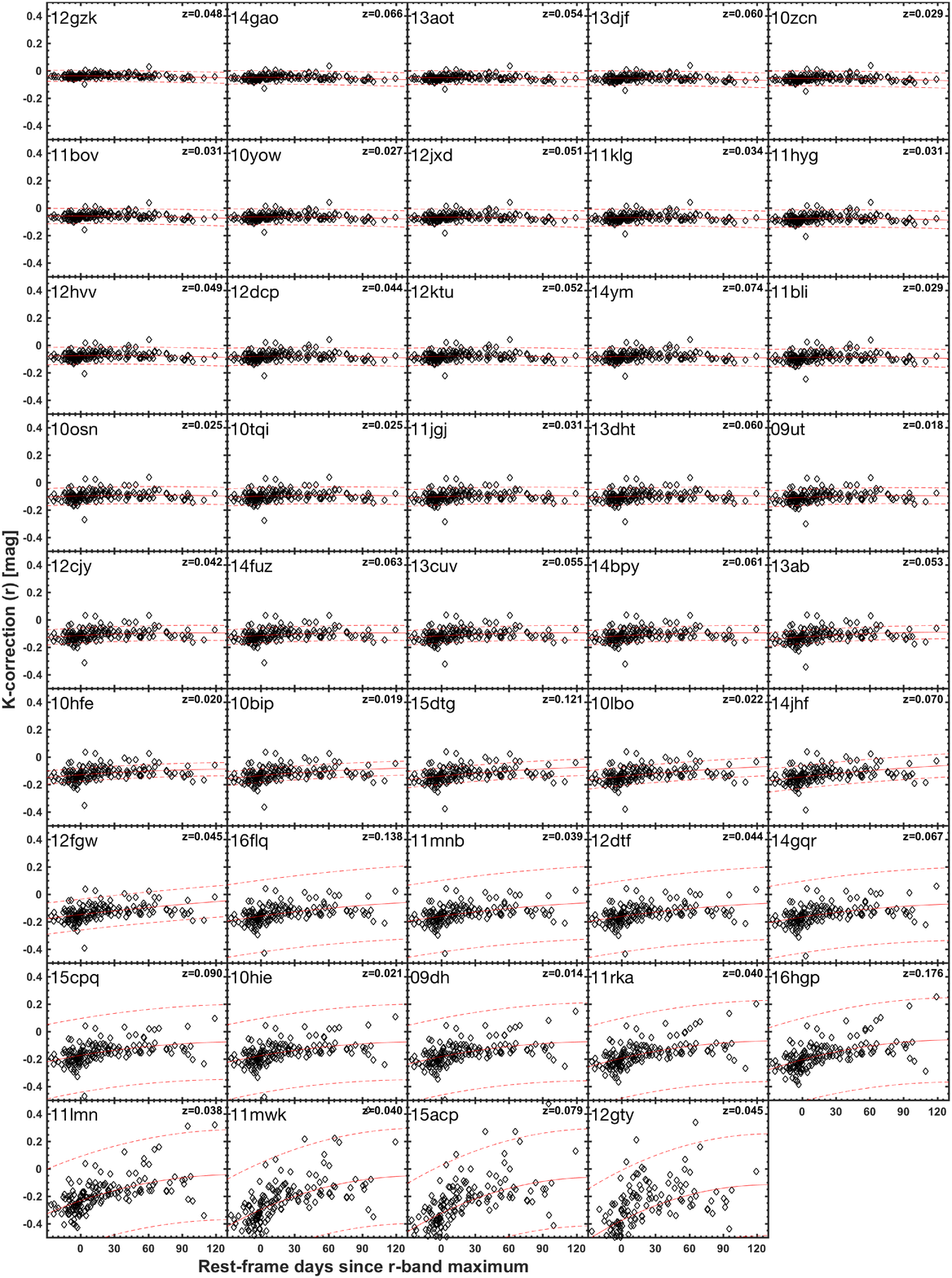}\\
 	\includegraphics[width=10cm]{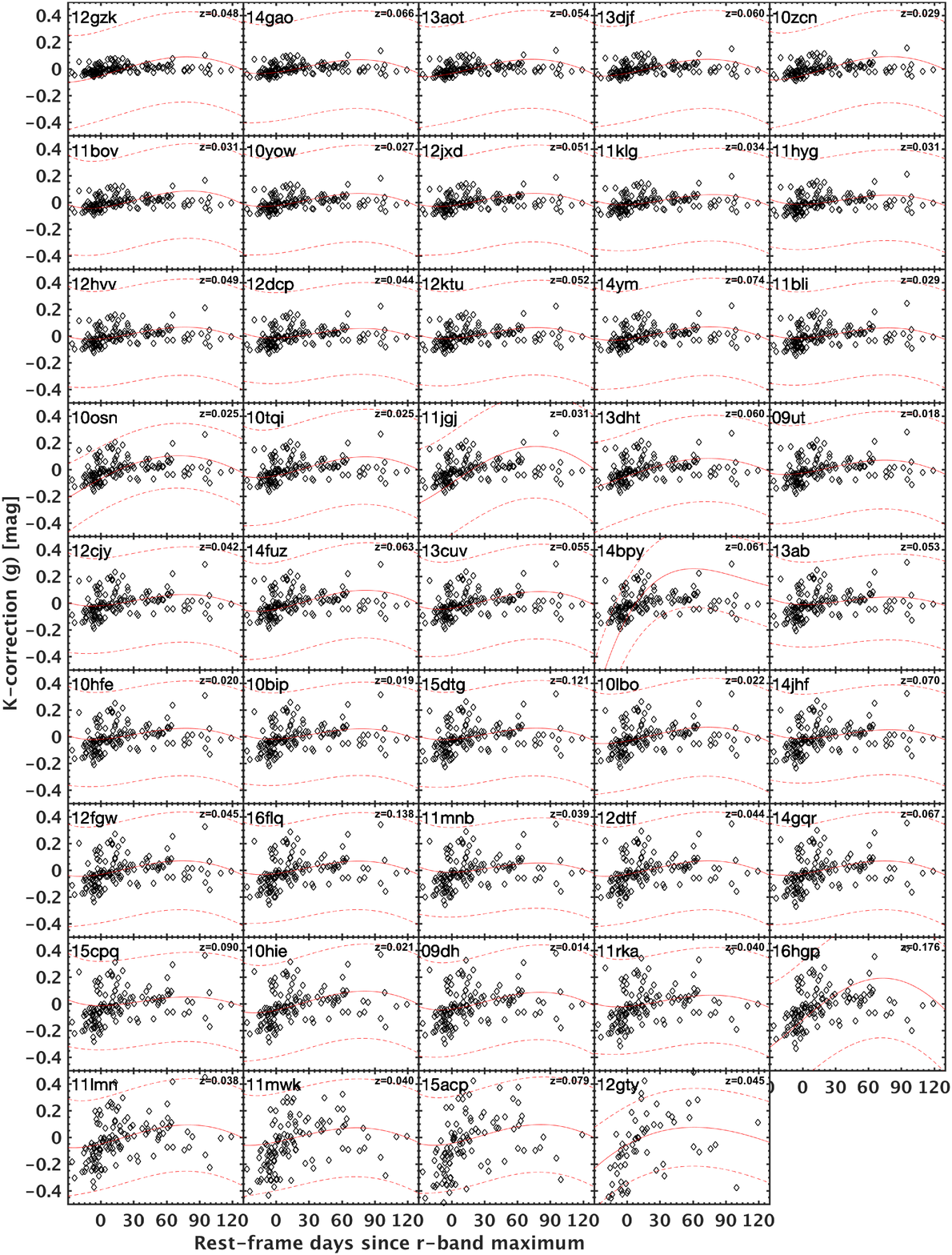}\\
 	\caption{K-corrections in the $r$ band for our SN sample. The solid red line represents the  second order polynomial fit, whereas the red dashed lines show the 1 $\sigma$ uncertainties. The SNe have been ordered according to increasing redshift.}
 	\label{Fig:Kcorr_r}
 \end{figure*}

When the peak was observed also (15 objects) or only (4 objects)\footnote{This was done for 19 SNe
namely PTF09dh, PTF11bli, PTF11bov, PTF11hyg, PTF11jgj, PTF11klg, PTF11lmn, PTF11mnb, iPTF12gzk, iPTF12hvv, iPTF14bpy, iPTF14fuz, iPTF14gao, iPTF14gqr, iPTF15acp, iPTF15cpq, iPTF15dtg, iPTF16flq and iPTF16hgp.} in the $g$ band, we performed the same fit in this band.
The observed phases were corrected by a factor of $(1 + z)$ to account for time dilation in order to obtain the phase in the rest frame.
The measured $t_{r}^{max}$ was used to determine the rest-frame phase of the SN spectra in our sample.  The spectra were used to compute average K-corrections for the $Bgri$ bands as a function of redshift and time since $t_{r}^{max}$.
The method we performed has been presented in \citet{Taddia19b} and makes use of all the available spectra of the sample to estimate the K-corrections for every single SN. The main reason to apply this method is the lack of a complete spectral follow-up for most of the SNe of the sample.
All the obtained K-corrections were then plotted  as a function of the phase and fitted with a second-order polynomial. These fits in the $r$ band are shown in Fig.~\ref{Fig:Kcorr_r}.

Overall, we can see that the K-corrections are negligible for most of the objects but when present, at higher z, they are more important at early epochs. We K-corrected all the $g-$ and $r-$band light curves interpolating the above mentioned polynomials at all epochs of the light curve observations. In the following analysis, we will always refer to our K-corrected and time-dilation corrected light curves.

\subsection{Light curve shape}\label{Sect:LCshape}
 
\begin{figure*}
	\centering
	\includegraphics[width=17cm]{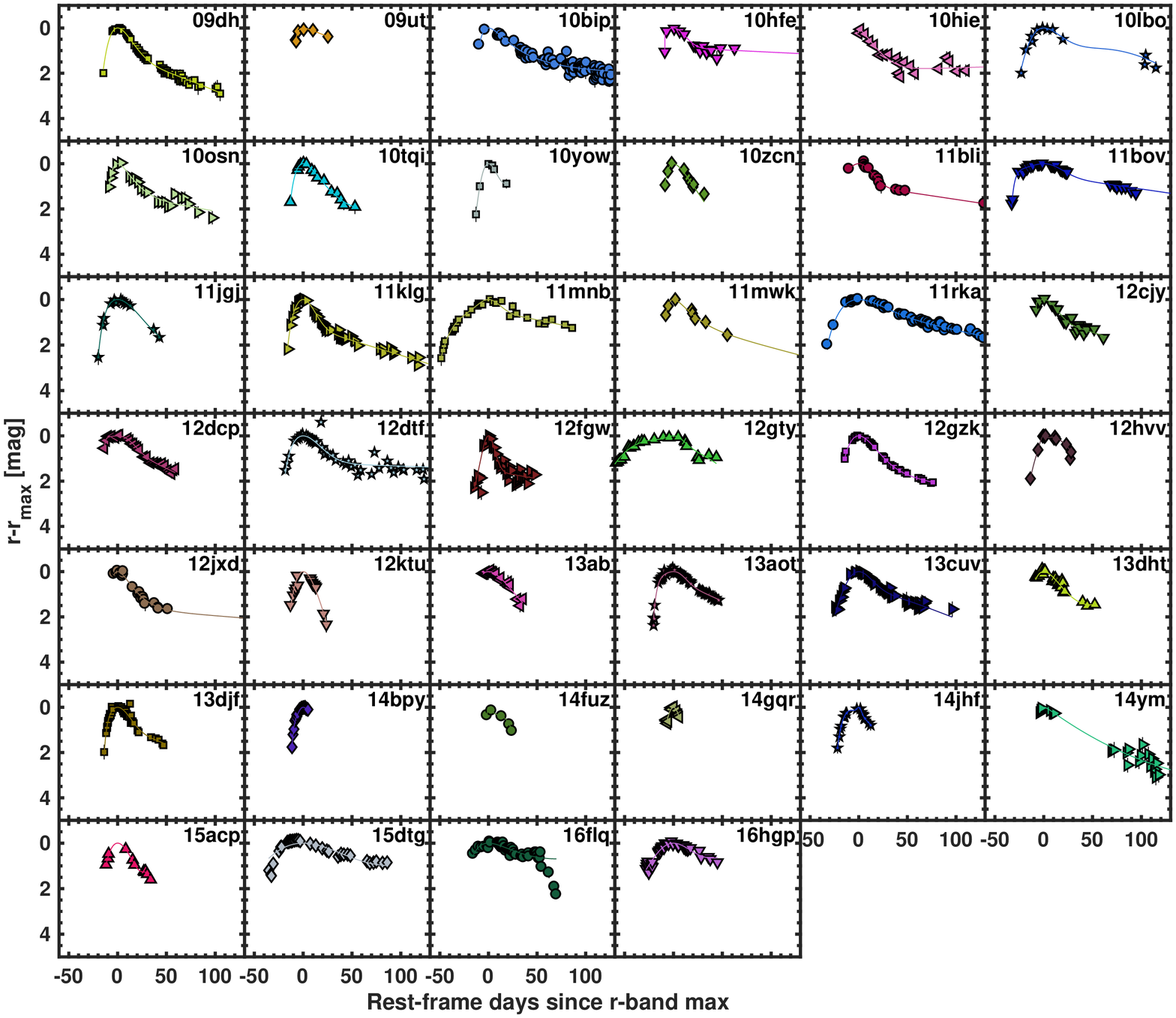}\\
	\includegraphics[width=17cm]{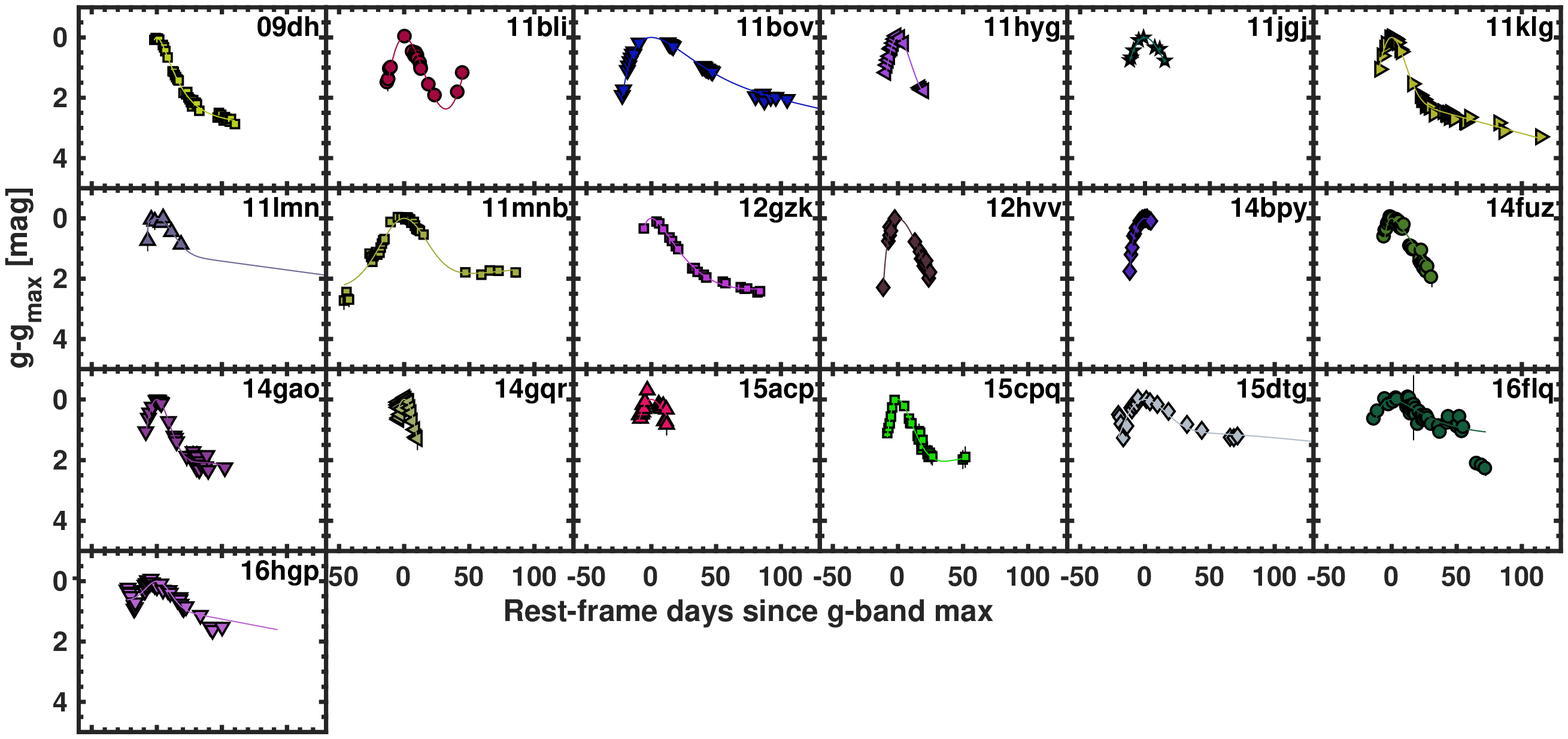}
	\caption{ \texttt{Upper panel:} Individual 40 SNe and their Contardo fits in the $r$ band.
	\texttt{Lower panel:} Individual 19 SNe and their Contardo fits in the $g$ band.}
	\label{Fig:LC_contardo}
\end{figure*}

We fitted the K-corrected $r$- and $g$-band light curves with the function provided by \citet{Contardo00} to characterise their shapes. This function includes an exponential rise, two Gaussian peaks, and a linear late decline. We included only the first of the two Gaussian peaks in our fit.
From this fit it is possible to derive the peak epoch and magnitude, the rise parameter $\Delta m_{-10}$ as well as the decline parameters $\Delta m_{15}$,  $\Delta m_{40}$ and the late linear decline slope. The parameter $\Delta m_{-10}$ measures how many magnitudes the light curve rises during the 10 rest frame days before peak. $\Delta m_{15}$ instead represents the decrease in magnitude 15 days after peak and $\Delta m_{40}$ at 40 days after peak.
The results of our fits to the $r$- and $g$-band light curves are shown in Fig.~\ref{Fig:LC_contardo}. 

In the top panel, each SN is represented individually in the $r$ band while the bottom panel shows the SNe in the $g$ band.

\begin{figure*}
	\centering
	\includegraphics[width=18cm]{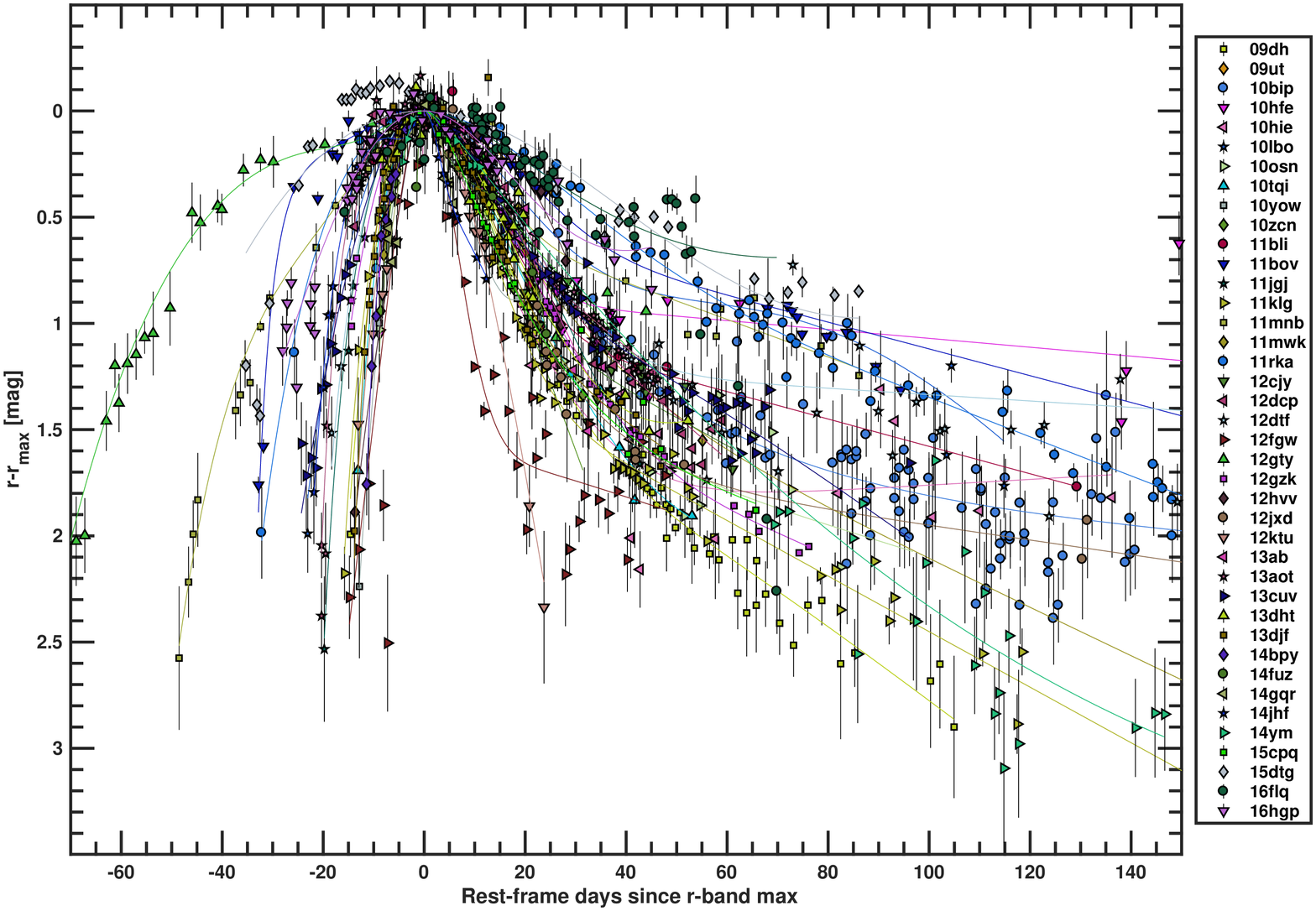}
	\caption{These are the $r$/band light curves for our 40 SNe Ic plotted together. The best Contardo fits are included as full lines. In this and following plots the individual SNe are represented with symbols and colors as provided in the legend to the right. The light curves are normalised at peak in order to illustrate their diversity.}
		\label{Fig:LC_contardo_all_r}
	\end{figure*}

We focus our analysis on the shape of the light curves in the $r$ band.
This leads to the exclusion of PTF11hyg, PTF11lmn, iPTF14gao and iPTF15cpq from our analysis since they only show the peak in the $g$ band.
In Fig.~\ref{Fig:LC_contardo_all_r} we present all our 40 SNe together to show the general shape of their light curves in the $r$ band.

This highlights the variety of rise and decline rates of the SNe of our sample.
Through a Monte-Carlo procedure and simulating N=100 light curves, we estimate the uncertainties on each of the light curve fit parameters according to their photometric uncertainties. The uncertainty on each parameter is represented by the standard deviations of the best fit parameters. These parameters and their estimated uncertainties are reported in Table~\ref{Table:LCshape}.

\begin{figure}
	\centering
	\includegraphics[width=9cm]{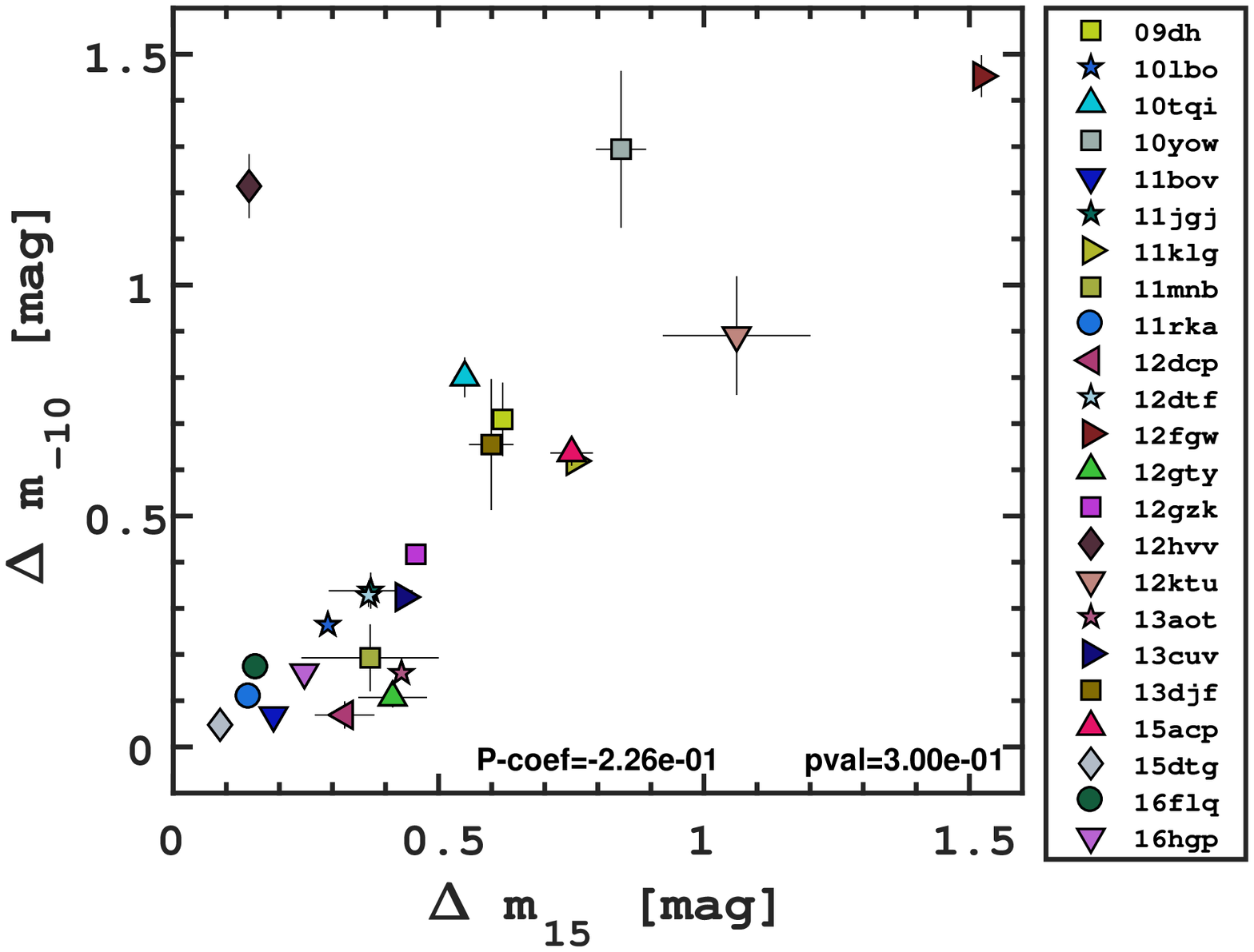}\\
	\includegraphics[width=9cm]{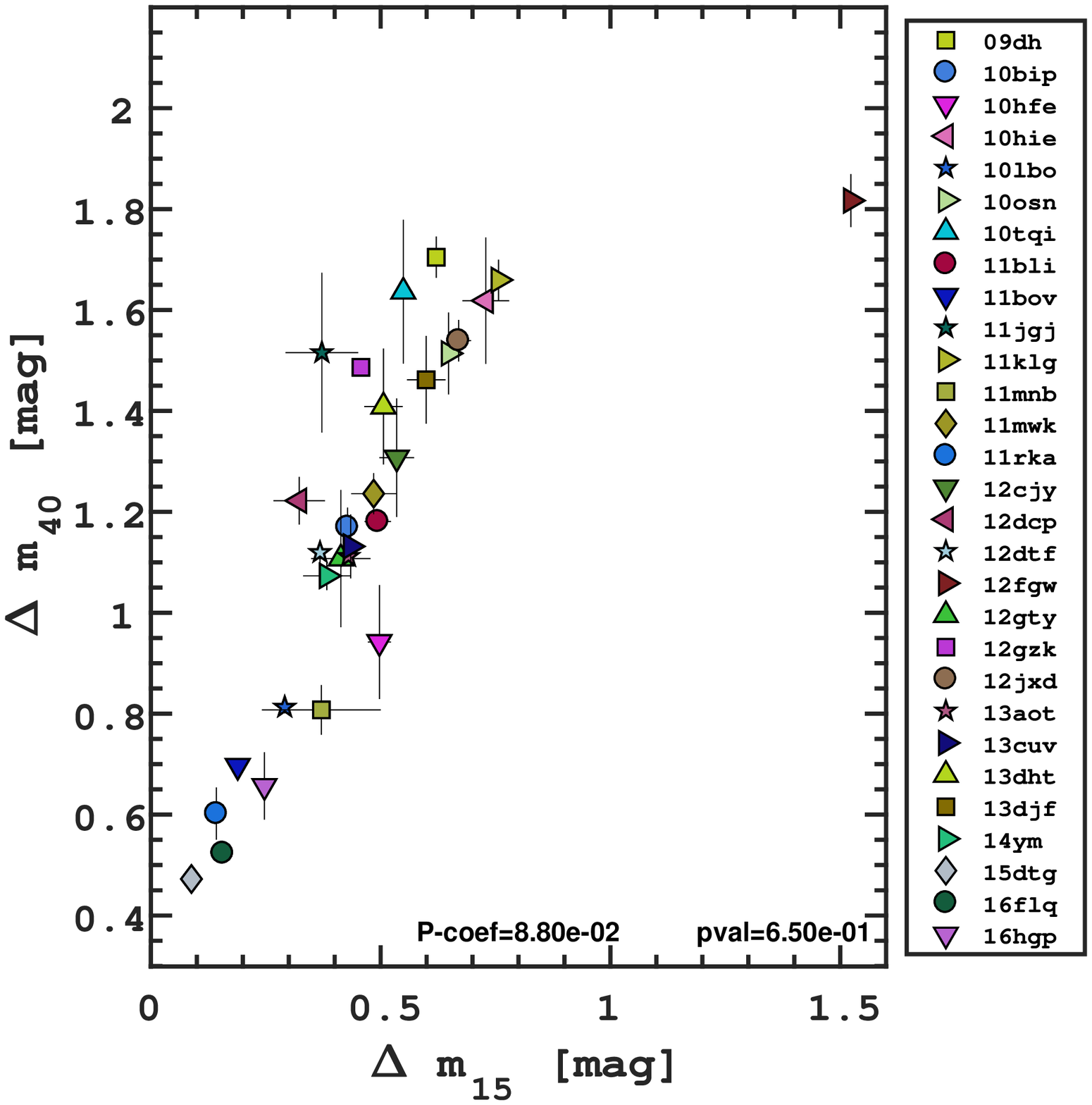}\\
	\includegraphics[width=9cm]{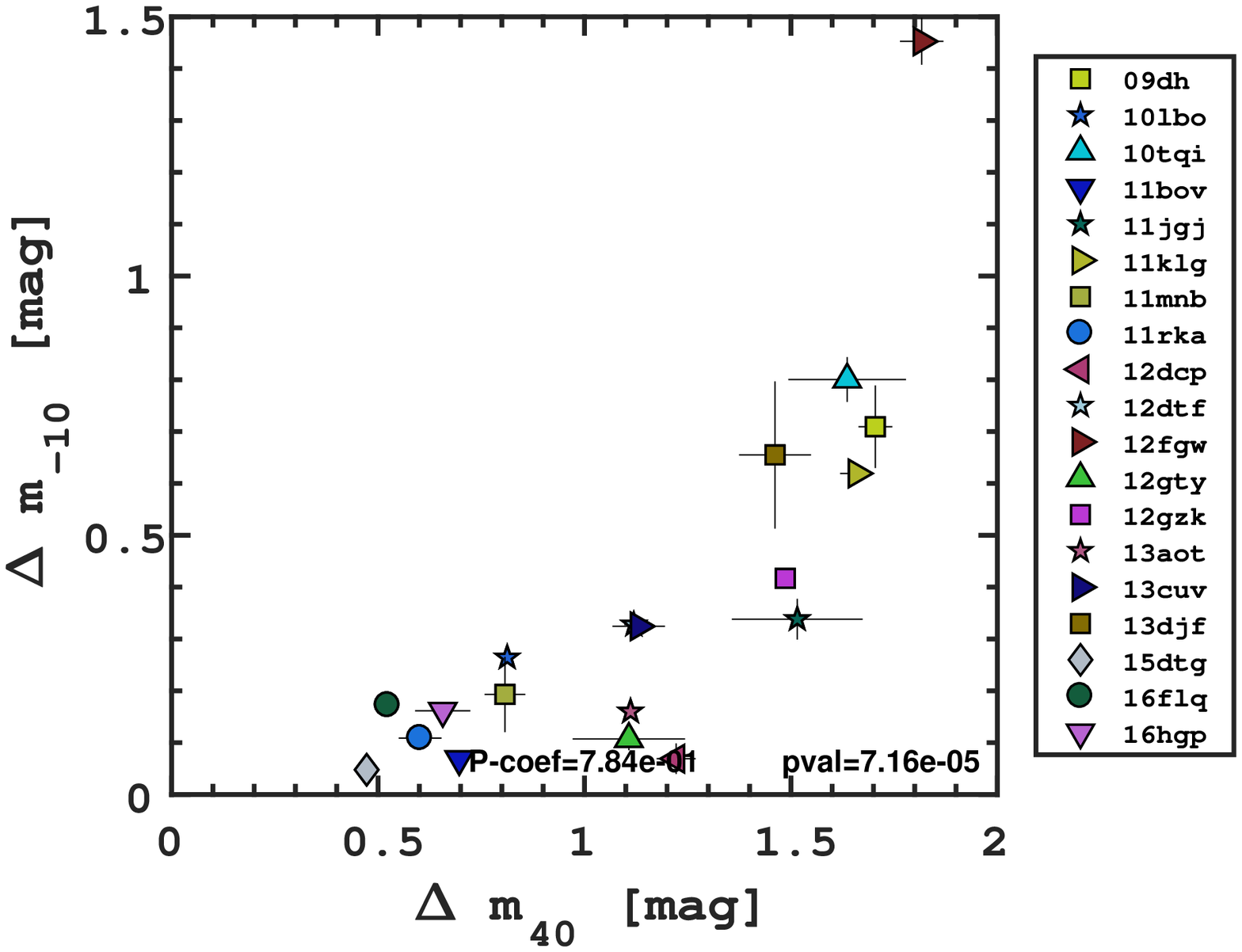}
	\caption{Correlations between rise and decay in the $r$ band. \texttt{Upper panel:} The plot shows $\Delta m_{15}$ against $\Delta m_{-10}$. 
	\texttt{Mid panel:}  $\Delta m_{15}$ versus $\Delta m_{40}$.  \texttt{Lower panel:} The plot shows $\Delta m_{40}$ against $\Delta m_{-10}$.Plots show a correlation among parameters and their p$-$values, along with the Pearson coefficients} are also reported.
	\label{Fig:LC_deltas}
\end{figure}

Figure~\ref{Fig:LC_deltas} suggests a correlation between $\Delta m_{15}$ and $\Delta m_{-10}$, with fast rising SNe also being fast declining.  We also show the relation between $\Delta m_{15}$ and $\Delta m_{40}$ and the one between $\Delta m_{-10}$ and $\Delta m_{40}$ which also seem to suggest a correlation among these parameters.
The $\Delta m_{-10}$ vs $\Delta m_{40}$ gives a similar relation as $\Delta m_{-10}$ vs $\Delta m_{15}$, 
implying that the fast rising SNe are the fast declining SNe also at 40 days past peak. The $\Delta m_{15}$ vs $\Delta m_{40}$ demonstrates that the SNe that decline fast during the first 2 weeks are also the ones that declined more rapidly at later phases.
We performed a Pearson test to quantify these correlations and the p-values are reported in Fig.~\ref{Fig:LC_deltas}. This test will be performed for all correlations in this work and p-values will be displayed in the plots. The $\Delta m_{40}$ and $\Delta m_{-10}$ show a lower p-value than the other two which indicates a stronger correlation, this is most likely due to the presence of an outlier in the other correlations.
We estimated the epoch of the peak in $g$ band for 19 SNe of the sample and compared it with the same estimate for the $r$ band. We find on average a shift between the peaks at $g$ and $r$ band of $4 \pm 2 \:  \mathrm{days}$, with the SN peaking in the $g$ band first. This is consistent with the estimate of $\sim 3 \:  \mathrm{days}$ presented in \citet{Taddia15}.
Armed with the average time shift between peaks in $g$ and $r$ band we can provide an estimate of the $r$-band peak for the 4 SNe which show a peak only in the $g$ band. These values are also presented in Table~\ref{Table:LCshape}.
 
 \subsection{Colours and host extinction}\label{Sect:ColHost}
 
\begin{figure*}
	\centering
	\includegraphics[scale=0.55]{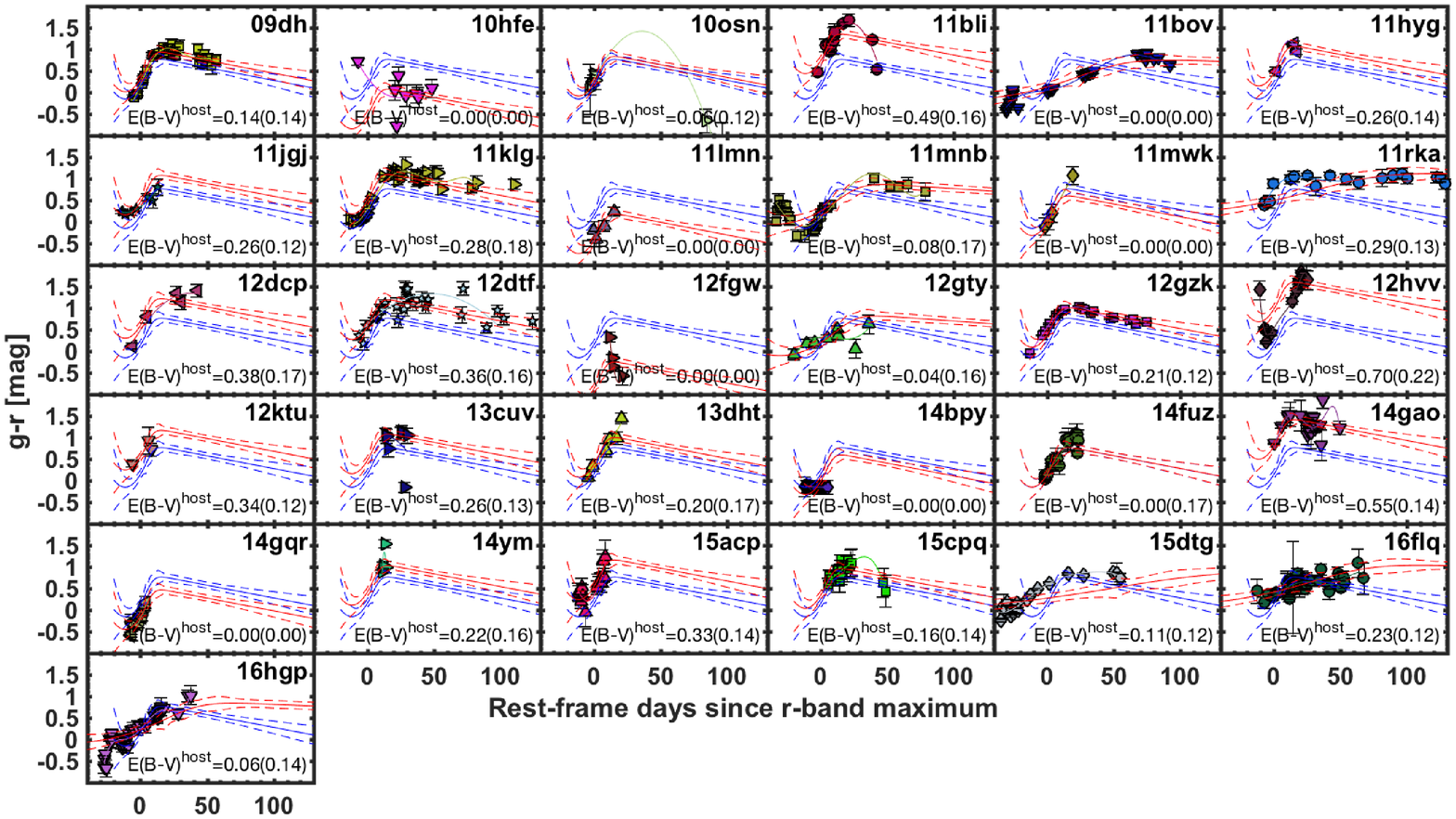}\\
	\includegraphics[scale=0.55]{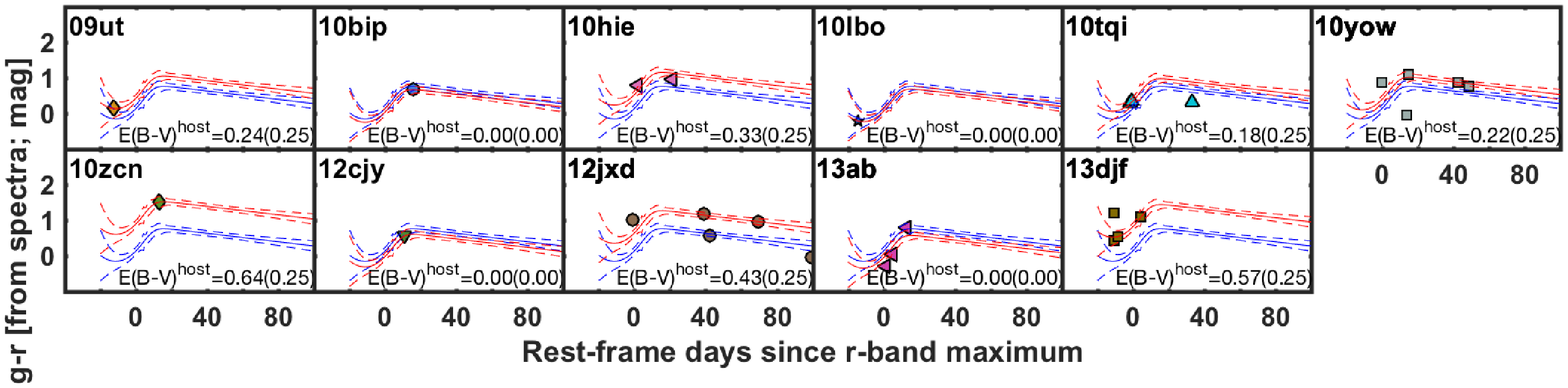}
	\caption{\texttt{Upper panel:} Individual (MW corrected) $g-r$ colour evolution for 31 SNe from photometry with the polynomial fits represented as solid lines. The red lines represent the fit of the data with the $g-r$ template. The blue line is the template for a Type Ic with no extinction from \citet{Stritzinger17}, and each panel shows the reddening in E(B$-$V) required to shift the colour curve to the data. This represents the estimated host extinction measured in magnitudes. \texttt{Bottom panel:} Same as above for the  11 SNe where the colours are calculated from spectroscopy.}
	\label{Fig:gmr}
\end{figure*}

We proceed to compute the colour evolution of the SNe, see Fig.~\ref{Fig:gmr}.

For 31 SNe with $g$- and $r$-band data available, we corrected for the Milky Way (MW) extinction adopting the MW $E(B-V)$ given in Table~\ref{Table:SNprop}, assuming $R_{V} = 3.1$ and a \cite{Fitzpatrick99} reddening law. We then interpolated the $r$ band to the $g$-band epochs to obtain the $g-r$ evolution.
The colour evolution tends to show an initial rise until approximately 20 days after peak, then it starts a shallow decline (getting bluer)
at later epochs (Fig. \ref{Fig:gmr}).
We estimated the host extinction from spectroscopy from the measurement of the equivalent width of the narrow Na I D absorption line.
We followed \citet[][their Eq.~1]{Taubenberger06} to get $E(B-V)$ and adopt an uncertainty of $\Delta E(B-V) = \: 0.2 \: \mathrm{mag}$.

We also estimated $E(B-V)$ using  \citet[][their Eq.~9]{Poznanski12} and notice that the values we get are in agreement with the previous estimates. The average difference between the two methods is $0.05 \: \pm \: 0.11$.  We note that for PTF11jgj and iPTF16flq we find the biggest difference for the extinction estimates from Na I D which also differ significantly from the estimates we get from the $g-r$ method. If we exclude these two SNe from the comparison we get that the average difference between the two methods is $0.03 \: \pm \: 0.06$.
Since the difference is rather small, we will adopt the $E(B-V)$ from \citet{Taubenberger06} when referring to extinction from Na I D through the manuscript.

In addition, we used the SN $g-r$ colours to estimate the host extinction, following the method described by \cite{Stritzinger17}.
We fit the $g-r$ colours of all the SNe with low-order polynomials, shown as solid lines in Fig.~\ref{Fig:gmr}. We then estimated the average $E(g-r)$ for each SN in the range between 0 and 20 days after the peak by computing the average difference between the fit of the observed $g-r$ and the assumed intrinsic $g-r$ colour. We adopted the $g-r$ template presented in \cite{Stritzinger17}.

The 6 SNe that present broader light curves as mentioned in Sect. \ref{Sect:sample} require a stretching factor to apply to our $g-r$ template in order to get a good fit (see Karamehmetoglu et al. in prep).
We estimated the stretch factor from the ratio between the average $\Delta m_{15}$ of the sample and the $\Delta m_{15}$ estimated for the individual SN with broad light curve, and then set the time scale from the ratio of the epochs of the peak in the $g-r$ colour evolution of one broad SN (iPTF16hgp) and one normal (PTF12gzk).
We then converted $E(g-r)$ into $E(B -V)_\mathrm{host}$, assuming $R_{V} = 3.1$.
The uncertainty of the $g-r$ template is included in the uncertainties of the host galaxy extinctions. The uncertainty also takes into account the standard deviation due to the difference between the epochs of the measured colour and the intrinsic $g-r$. We also used the spectra corrected for redshift and Milky Way extinction to build  $g-r$ color curves for some SNe\footnote{In particular, spectra were used for 11 SNe, namely PTF09ut, PTF10bip, PTF10hie, PTF10lbo, PTF10tqi, PTF10yow, PTF10zcn, PTF12cjy, PTF12jxd, iPTF13ab and iPTF13djf}
for which it was not possible to use the photometry (see bottom panel of Fig.~\ref{Fig:gmr}). 
The computed $E(B-V)_\mathrm{host}$ values are reported in Table~\ref{Table:SNprop}.
We note that for two SNe\footnote{This has been adopted for SNe iPTF13aot and iPTF14jhf.} it was not possible to get a $g-r$ evolution.

\begin{figure}
	\includegraphics[width=9cm]{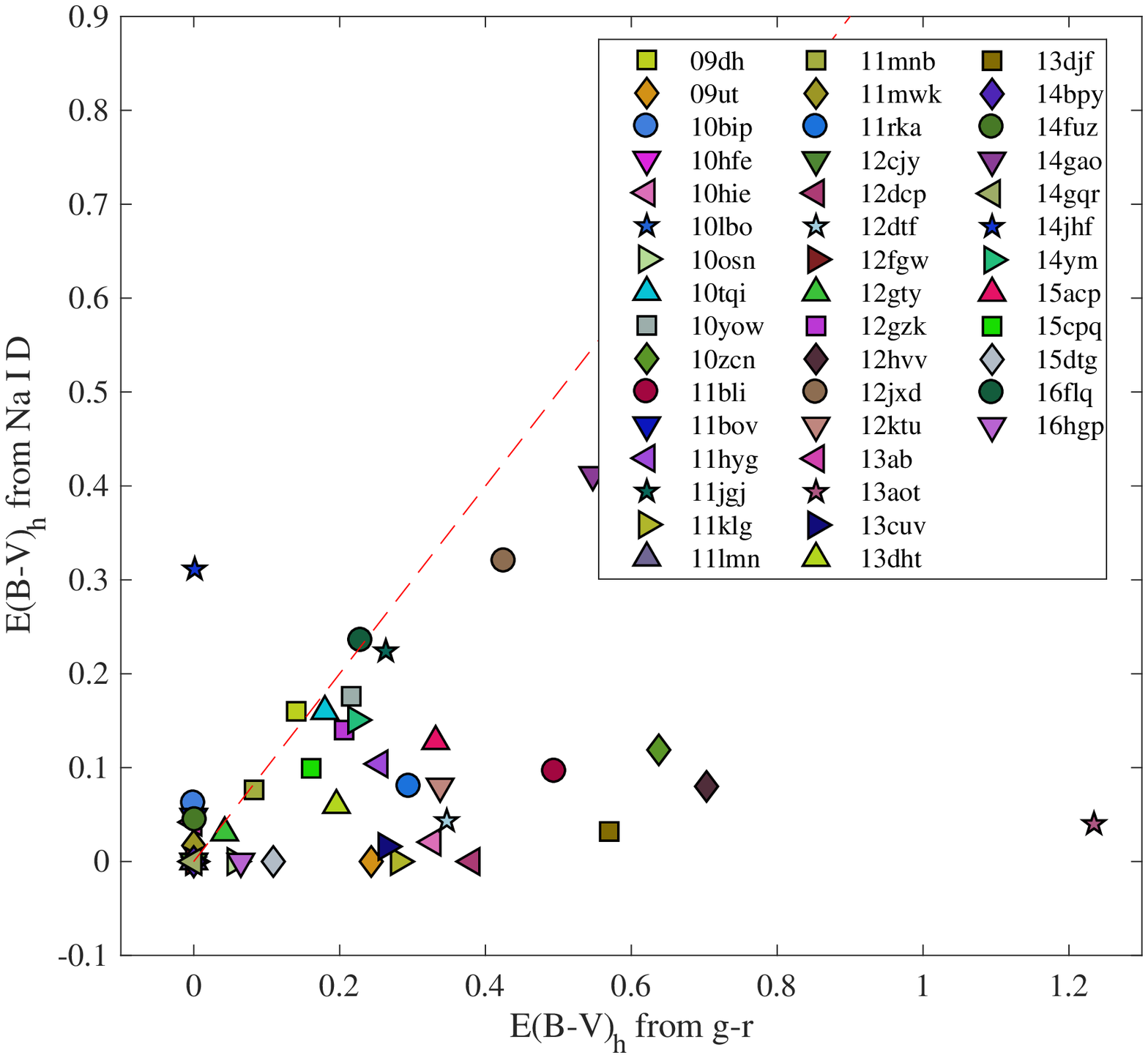}
	\caption{A comparison between the extinction estimated from the Na I D absorption versus that estimated from the $g-r$ colour evolution. The red line represents a one-to-one relation, but clearly the Na I D gives consistently lower estimates. Units are in magnitudes.}
	\label{Fig:ebmv_vs_ebmv}
\end{figure}

When we compare the host galaxy extinction estimates obtained from the SN colour comparison to that obtained from the Na I D absorption lines, we notice that the first seems to provide overall higher values (Fig.~\ref{Fig:ebmv_vs_ebmv}). 

Both of these methods come with considerable uncertainties and assumptions. We note
that some of our spectra do not have enough signal-to-noise (S/N) close to the Na I D line to properly detect it. On the other hand, the $g-r$ method relies on an intrinsic colour curve template and on assuming homogeneity of the colour evolution for these SNe.
In this work we will adopt the extinction estimated from the Na I D, unless otherwise specified. The main reason is that we want to compare our results with those published in the literature, which most often have used this method.
However, throughout the analysis we will discuss how some values will be affected if we instead chose the extinction estimated from the second method.

 \subsection{Absolute magnitudes}
 
\begin{figure*}
	\includegraphics[width=16cm]{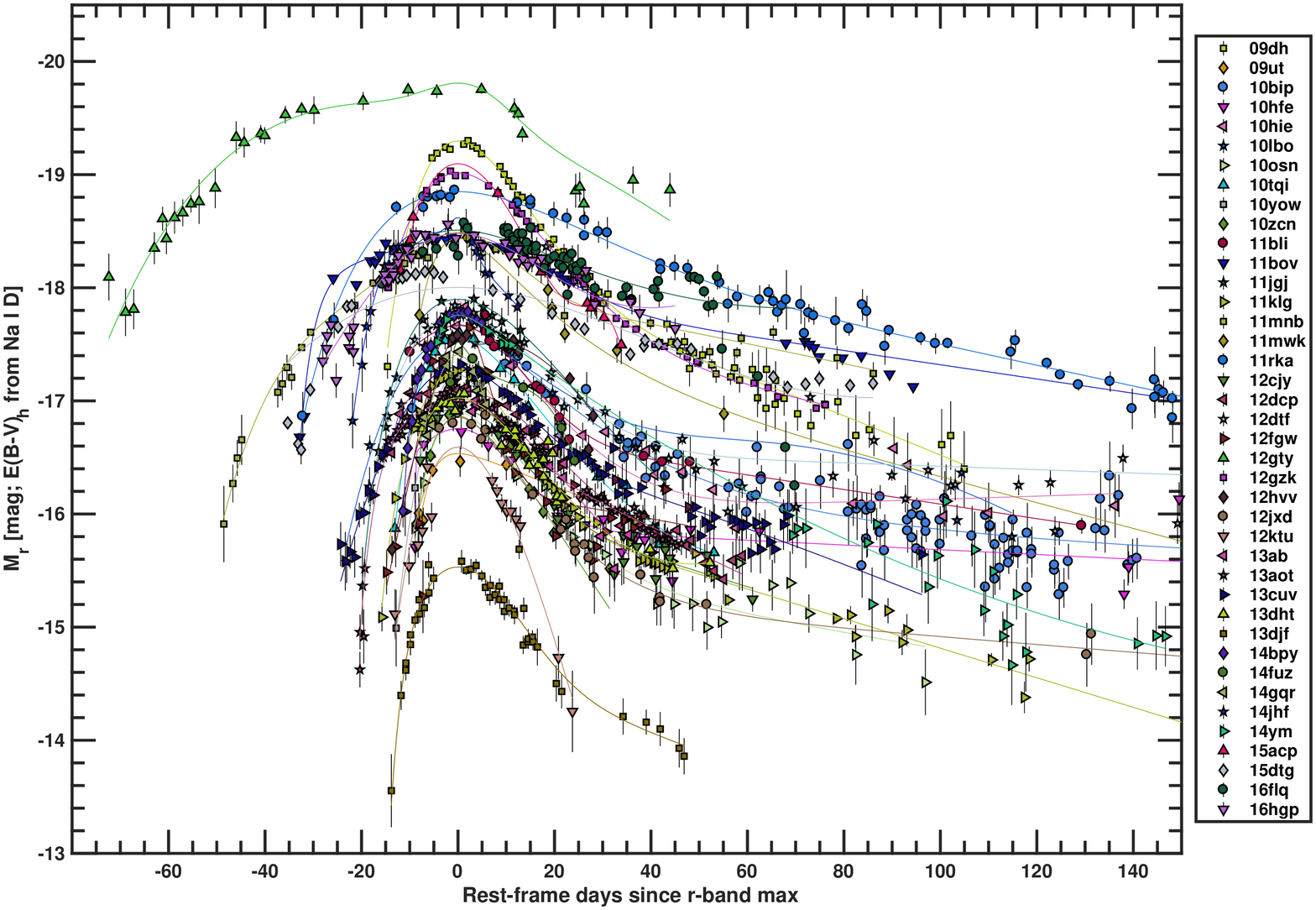}\\
	\includegraphics[width=16cm]{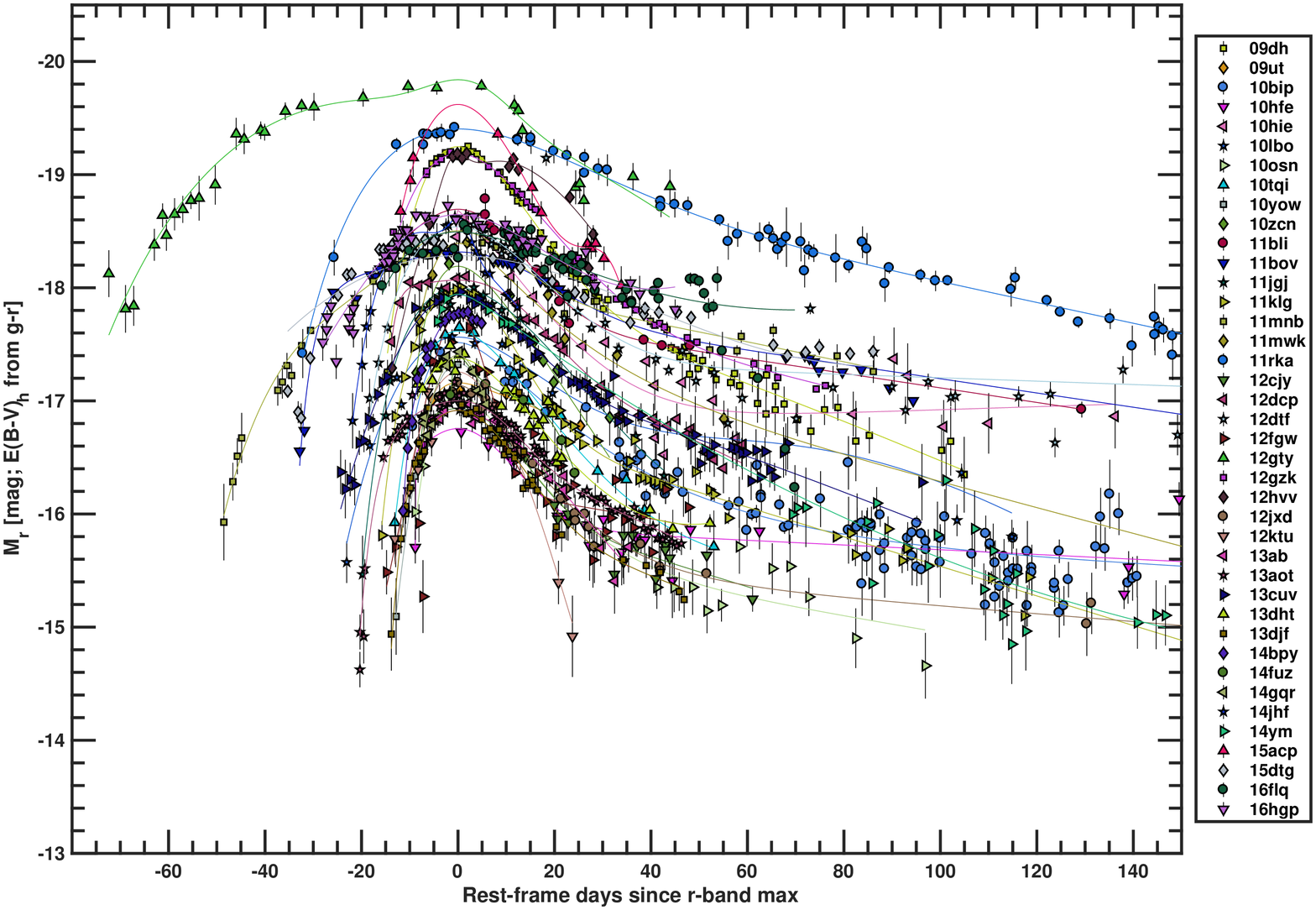}\\	
	\caption{
	\texttt{Upper panel:} Absolute magnitude in $r$ band of the 40 SNe of the sample when extinction is estimated from the Na I D absorption. \texttt{Bottom panel:} Absolute magnitude in $r$ band of the 40 SNe of the sample when extinction is estimated from $g-r$ colour evolution. For SNe iPTF13aot and iPTF14jhf we assumed the extinction from the Na I D in both cases, since there is no estimate from $g-r$.}
	\label{Fig:LCebmv_vs_ebmv}
\end{figure*}

We applied the presented corrections; Milky Way and host extinctions, distances and K-corrections, to the light curves to obtain the absolute magnitudes, 
see Fig.~\ref{Fig:LCebmv_vs_ebmv}.

\begin{figure}
	\includegraphics[width=9cm]{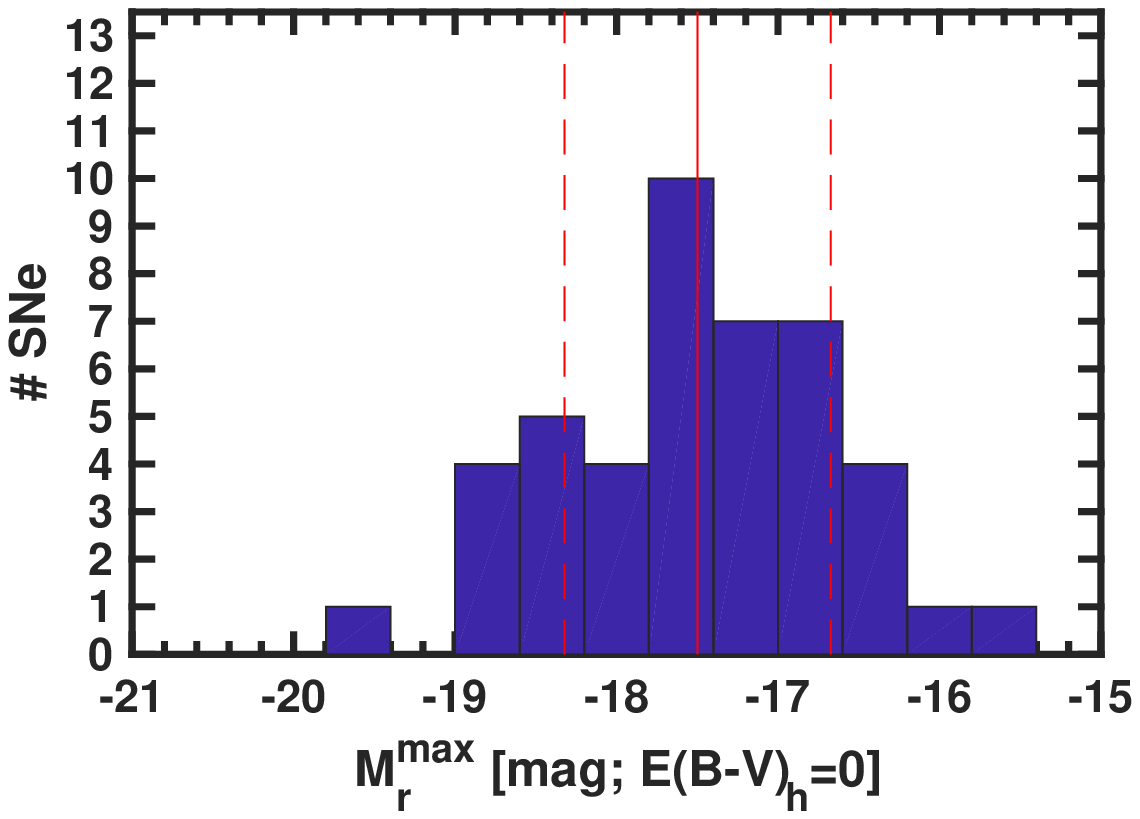}\\
		\includegraphics[width=9cm]{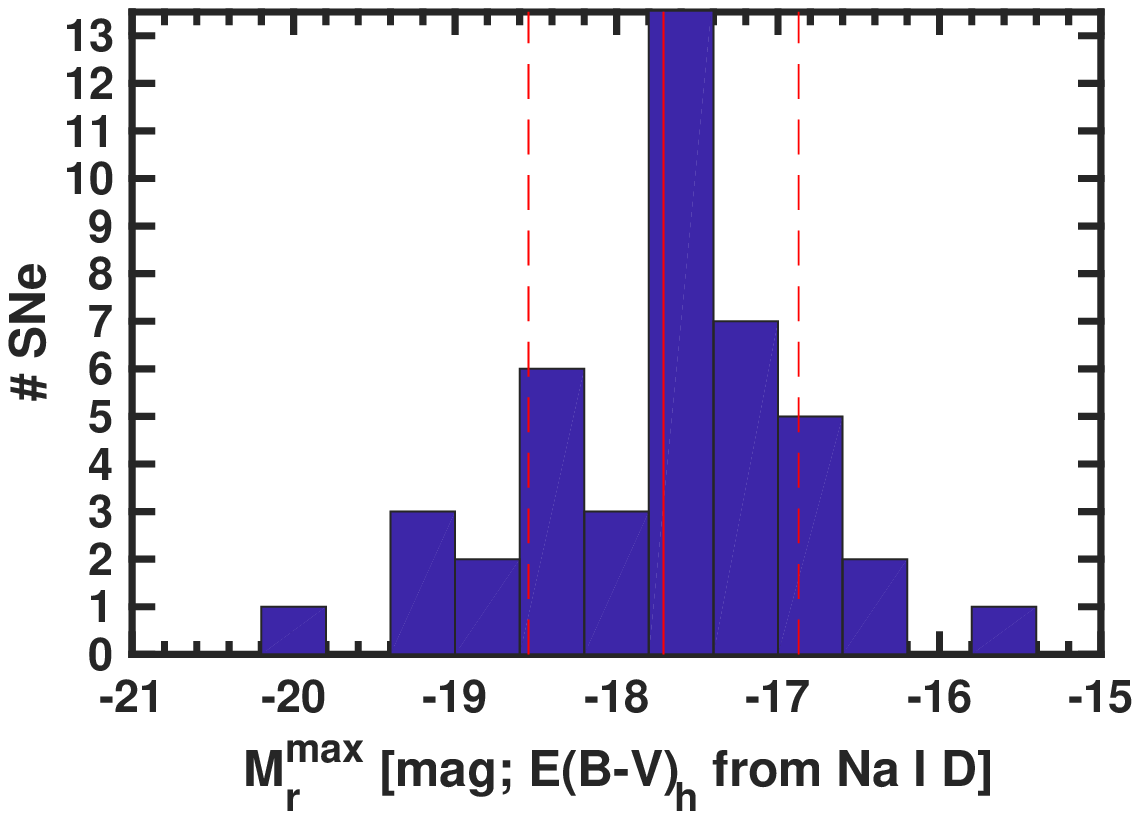}\\
		\includegraphics[width=9cm]{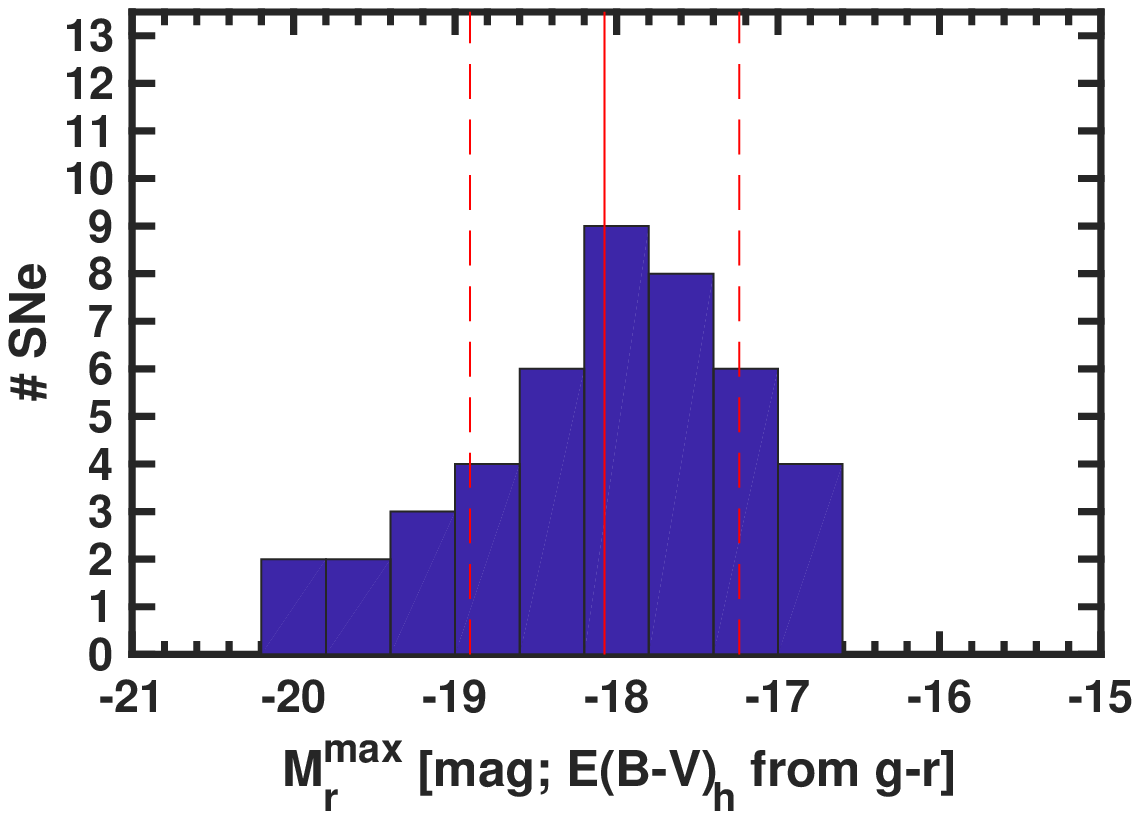}\\
	\caption{Histogram representation of the absolute magnitudes at peak in the $r$-band distribution of the sample. \texttt{Upper panel} Distribution obtained correcting for the distance and the MW extinction. \texttt{Middle panel} Distribution obtained including also the extinction from the host galaxy, estimated through the Na I D absorption.
	\texttt{Bottom panel} Distribution obtained including instead the host extinction from the colours.
		}
	\label{Fig:AbsmDist}
\end{figure}

The uncertainty on the absolute magnitudes takes into account the uncertainties due to the host extinction estimates and the photometric errors. In addition, the uncertainty on the distance adds a systematic error of $\pm \: 0.15 \: \mathrm{mag}$ which has not been included in the figure. This systematic is for an adopted uncertainty on $H_{0}$ of $\pm5$~km~s$^{-1}$, which dominates uncertainties from the peculiar velocities at the redshifts of our sample SNe.
The distribution of the $r$-band absolute magnitudes at peak is shown in Fig.~\ref{Fig:AbsmDist}.

\begin{figure}
	\centering
	\includegraphics[width=9cm]{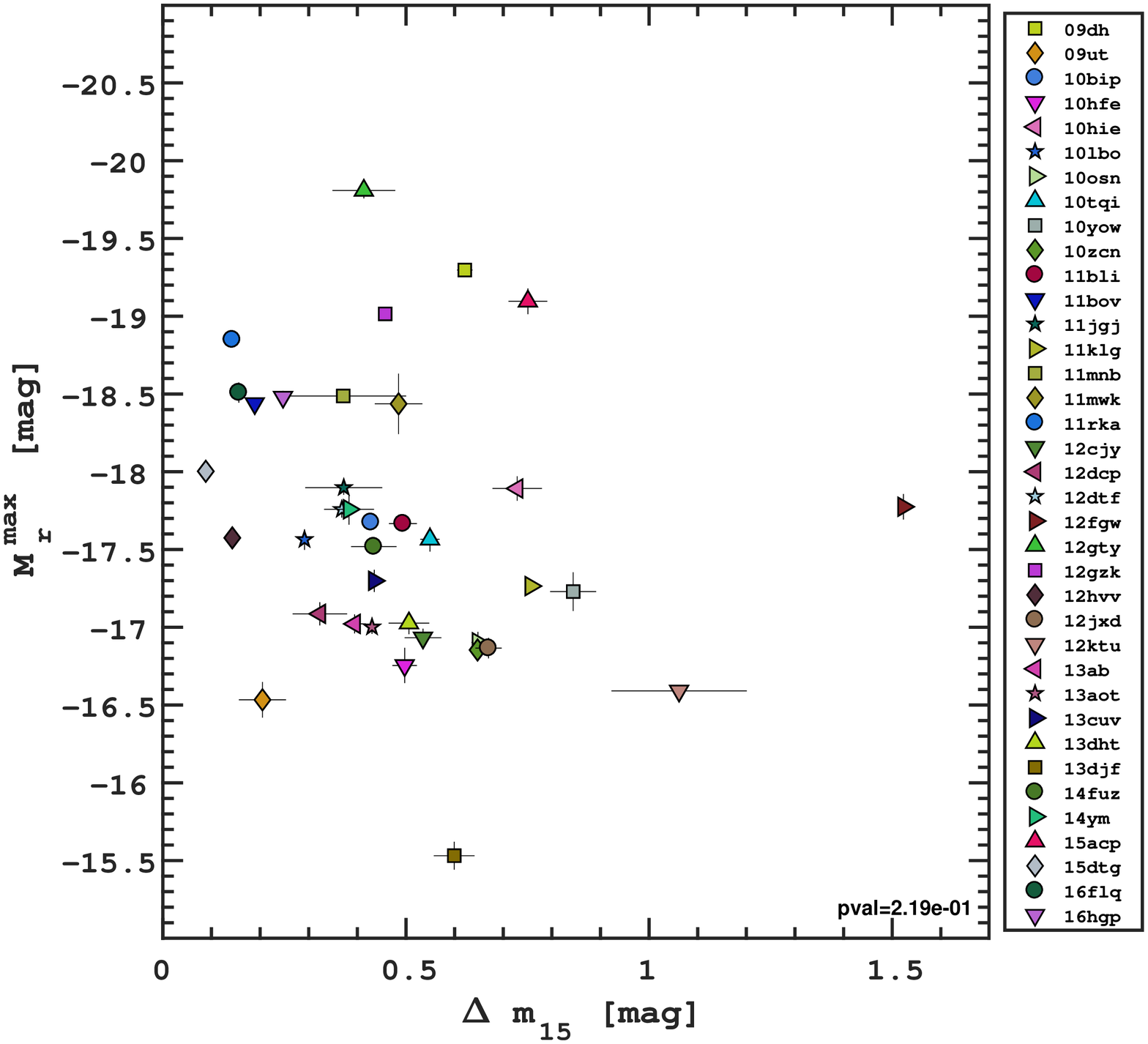}\\
	\caption{
	Peak absolute $r$-band magnitude vs $\Delta m_{15}$. The plot shows no correlation.
	}
	\label{Fig:absm_relation}
\end{figure}

Our $r$-band magnitudes span the interval $-15.45$ to $-19.73 \: \mathrm{mag}$ when the host extinction has not been accounted for, giving an average of $<M_{r}^{\mathrm{peak}}> = -17.50 \pm 0.82 \: \mathrm{mag}$.
It ranges from $-15.54$ to $-19.81 \: \mathrm{mag}$ when the host extinction from Na I D is included, with an average of $<M_{r}^{\mathrm{peak}}> = -17.71 \pm 0.85 \: \mathrm{mag}$.
If we instead consider the extinction estimates from $g-r$, the interval is $-16.91$ to $-19.84 \: \mathrm{mag}$ and an average of $<M_{r}^{\mathrm{peak}}> = -18.07 \pm 0.84 \: \mathrm{mag}$. All values for each SN in the sample are reported in 
Table~\ref{Table:Mag_r}. We notice how PTF12gty is the brightest SN in the sample with an absolute peak magnitude of $-19.81$.\\
The absolute magnitude ranges available in the literature are $M_{r}^{\rm{peak}} = -18.26 \pm 0.21 \: \mathrm{mag}$  \citep{Taddia15}; $M_{r}^{\mathrm{peak}} = -17.64 \pm 0.26 \: \mathrm{mag}$  \citep{Taddia18b} and  $M_{R}^{\mathrm{peak}} = -18.3 \pm 0.6 \: \mathrm{mag}$ \citep{Drout11}.
The average peak magnitude in the $r$ band estimated for our sample is thus in agreement with the ones from the literature.
We compared these values also with the (i)PTF sample of SNe Ic-BL \citep{Taddia19b} where the average peak magnitude is $-18.7 \pm 0.7 \: $ mag. Our SNe Ic are on average fainter than the SNe Ic-BL.
We investigated the absolute $r$-band magnitude peak versus $\Delta m_{15}(r)$ behaviour, to test if there is a Phillips-like relation as for 
SNe~Ia \citep{Phillips93}. We found that SNe Ic do not show any clear correlation (see Fig.~\ref{Fig:absm_relation}).

This is in agreement with previous studies on SE SNe (\citealt{Prentice16,Lyman16,Drout11}). Also a dedicated study on SNe Ic-BL has shown that there is no evidence for such a correlation \citep{Taddia19b}.\\
For SNe having data more than 70 days past peak, we also measured the slope at late epochs.
We investigated the $\Delta m_{15}(r)$ versus the slope and we did not see any clear correlation. This is not in agreement with what \citet{Taddia18b} found in their work.
Our $g$-band peak magnitudes for 19 SNe
span the interval $-15.86$ to $-18.91 \: \mathrm{mag}$ when the host extinction is not taken into account and it ranges from $-17.10$ to $-19.51 \: \mathrm{mag}$ when included, with an average value of $-17.99 \pm 0.69 \: \mathrm{mag}$. Finally, if we consider the extinction we get from the $g-r$ method the interval is $-17.04$ to $-19.44 \: \mathrm{mag}$. In this case the average is $-18.39 \: \pm 0.65 \: \mathrm{mag}$.
The average value for the peak magnitude in the g$-$band is in agreement with the $-17.28 \pm 0.24$ found by \citet{Taddia18b}.

\subsection{Explosion epochs and rise times}

\begin{figure*}
	\centering
	\includegraphics[width=18cm]{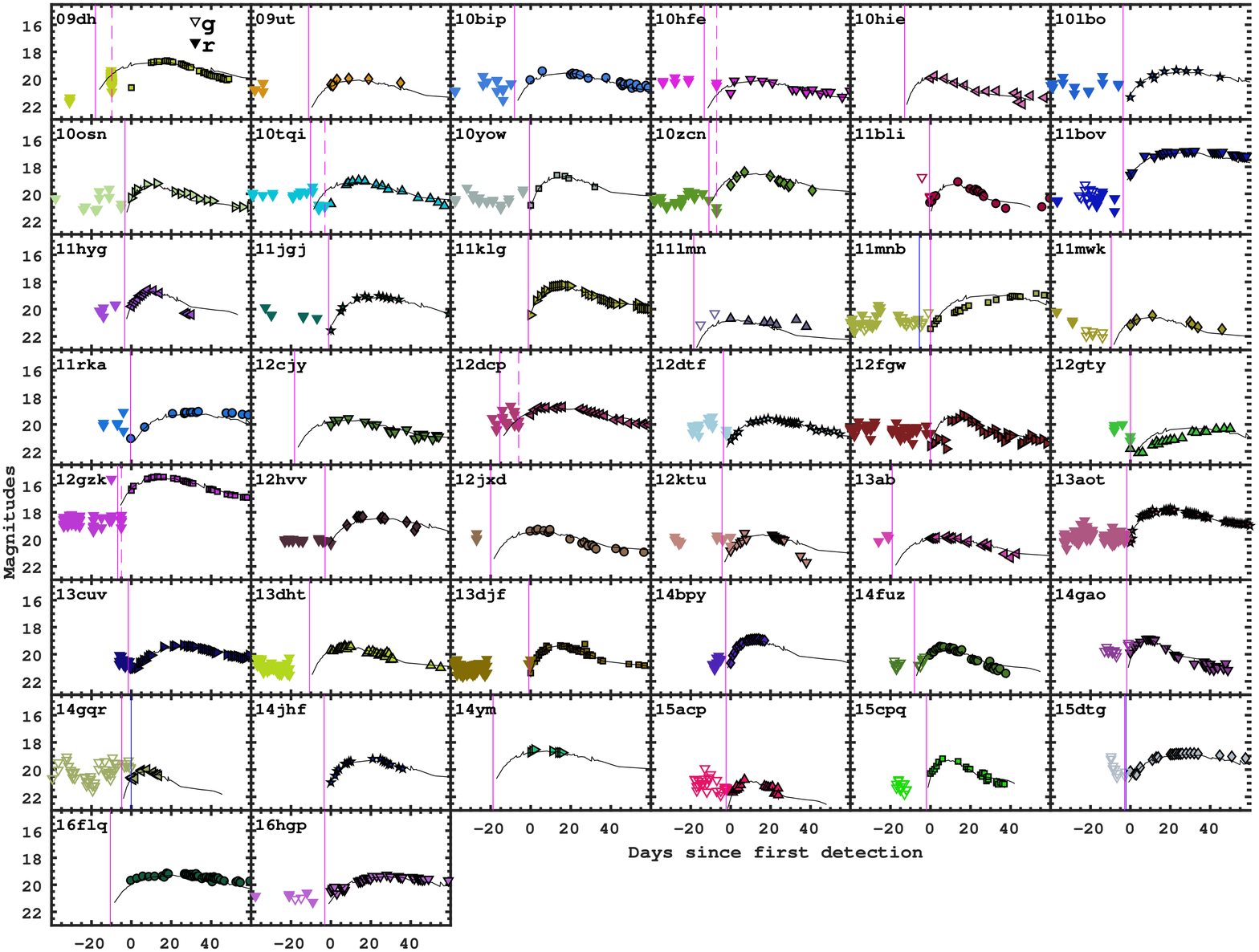}
	\caption{The plot shows the fit of the light curve of iPTF13djf to the other SNe of the sample to estimate the explosion epoch. The open symbols represents the pre-explosion limits. Solid magenta lines represent the explosion epochs estimated from the fit. Dashed magenta lines represent the last non-detection assumed as explosion epoch. Blue solid lines represent explosion epochs available from literature.}
	\label{Fig:texplo}
\end{figure*}

The separation between last 
non-detection and first detection for all the SNe of the sample varies in the interval $1-30$ days, with the exception of six SNe\footnote{PTF10hie, PTF11klg, iPTF12cjy, iPTF14jhf, iPTF14ym and iPTF16flq.} that do not have last non-detection within 50 days prior the first detection.
In order to estimate the explosion epochs for each SN, we compare their $r$-band light curves with the $r$-band light curve of iPTF13djf. This supernova has a good photometric coverage and well determined explosion epoch, with a limit on the discovery date of only $\pm$1 day. Since the explosion epoch of iPTF13djf is well constrained, as is the peak epoch in the $r$ band, we use it as a template and the stretch of the best fit allows us to infer the explosion epoch for all other SNe in the sample.
The light curve of iPTF13djf is stretched in time and shifted in magnitude to fit our SN light curves until +30 days post peak. The estimated explosion epochs were checked against the pre-explosion limits for consistency. We adopted $\pm 2~\: \mathrm{days}$ as a conservative estimate of the uncertainties on the explosion epochs.\\
In a few cases when this method did not give results consistent with the pre-explosion limits, we assume the last non-detection as the explosion epoch.\footnote{We assumed the last non detection as explosion epoch for PTF09dh, PTF10hfe, PTF10tqi, PTF10zcn and PTF12gzk.}
We note that in these cases the values we will estimate for the rise time will have to be considered as an upper limit.
When an estimate for the SN explosion epoch was available from literature, we adopted the latter as our explosion epoch. This was the case for PTF11mnb \citep{Taddia18a}, iPTF14gqr \citep{De18} and iPTF15dtg \citep{Taddia16}. The best fits and the obtained explosion epochs are shown in Fig.~\ref{Fig:texplo}.

The inferred explosion epochs are reported in Table~\ref{Table:texplo_trise}.
The explosion epochs and the epochs of the maximum in $r$ band allow us to compute the rest-frame $r$-band rise time, these are provided in Table~\ref{Table:texplo_trise}.
The average rise time we get is $25.3$ days which is somewhat higher than the $16.8$ days found by \citet{Lyman16} and the $13.3$ days found by \citet{Taddia18b}. This is most likely due to the fact we have more slow rising SNe than in their samples.
 
 \section{Construction of the Bolometric Light curves} \label{Sect:bolo}
 
Modeling of the bolometric light curves can help derive parameters on the supernova progenitors and on the explosion physics. 
To accomplish this, we need to estimate the explosion epochs and construct the bolometric light curves.

\subsection{Bolometric lightcurves}

\begin{figure*}
	\centering
	\includegraphics[width=18cm]{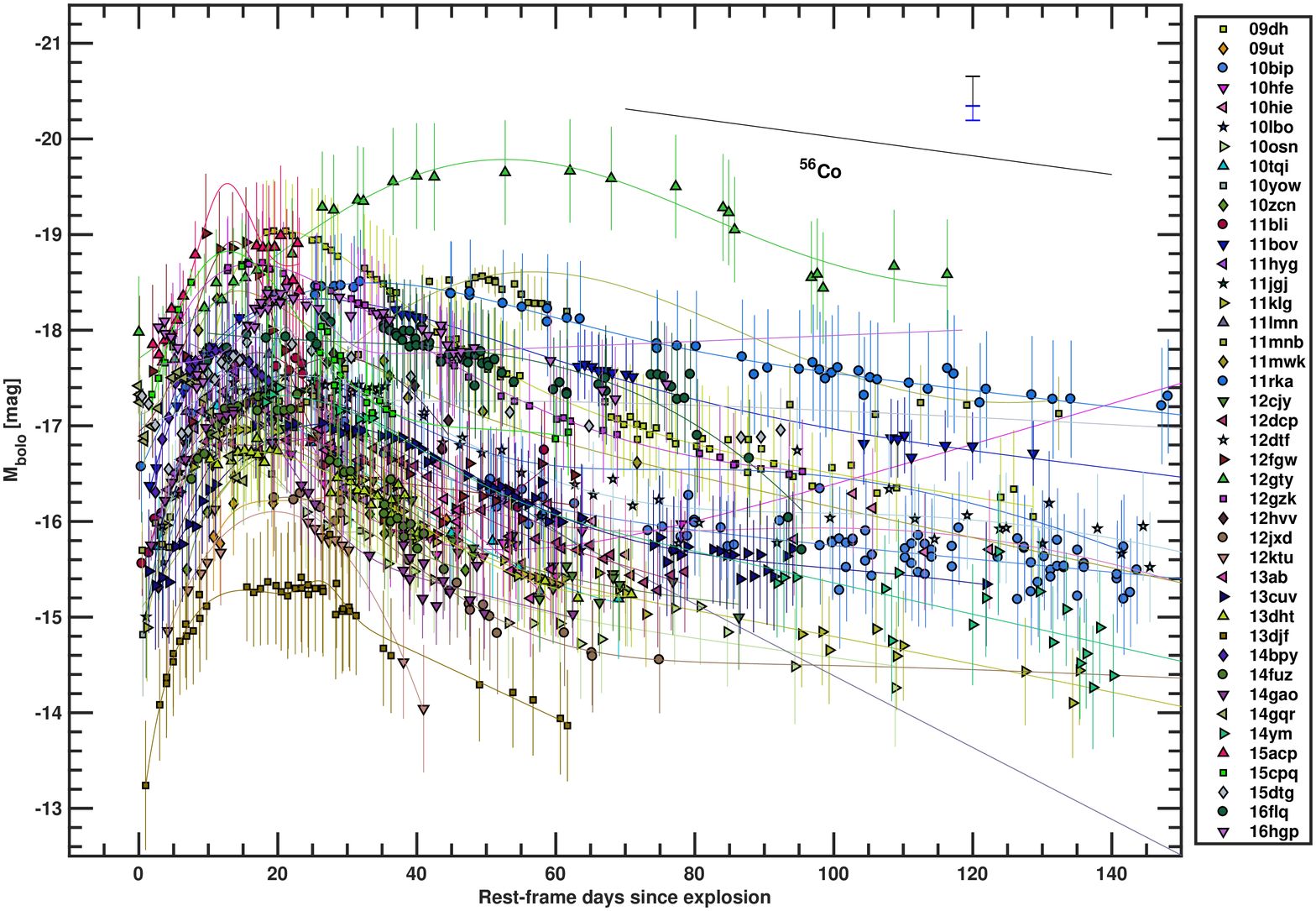}
	\caption{Bolometric light curves of 42 Type Ic SNe . The solid lines represent the Contardo fits performed on every individual light curve. The slope of the radioactive cobalt decay, 0.098 mag per day is illustrated in the upper right corner. There we also include a representative error bar that includes the uncertainty in distance, and extinction, respectively, which are not included in the errors on the data points.}
	\label{Fig:bolocontardo}
\end{figure*}

Due to the lack of a complete multiband coverage, in particular at early epochs, we used the absolute $r$-band light curves and the fit of the $g-r$ colour evolution to compute the bolometric light curves, making use of the bolometric corrections for SE SNe presented by \citet{Lyman14}. In this way we are able to create bolometric light curves covering all the phases.
The bolometric light curves of 
12 SNe\footnote{
PTF11bli, PTF11bov, PTF11hyg, PTF11lmn, PTF11mnb, iPTF14fuz, iPTF14gao, iPTF14gqr, iPTF15acp, iPTF15cpq, iPTF15dtg and iPTF16hgp.} were built by applying the bolometric correction directly to the $g$ band, which is
what is needed for the method of \citet{Lyman14}.
For the other 30 SNe, we interpolate the $g$ band from the $r$ band and then applied the bolometric correction.  Only for iPTF13aot and iPTF14jhf were we unable to  build a bolometric light curve due to a lack of $g-r$ evolution and these are therefore excluded from the analysis.
The final bolometric light curves as a function of days since explosion are shown in Fig.~\ref{Fig:bolocontardo}.

The systematic uncertainties due to the bolometric correction ($0.076~\: \mathrm{mag}$) and on the distance ($0.15~\: \mathrm{mag}$) are not included in the errors of each bolometric light curve. 

\subsection{Analysis of the bolometric light curve shape}

\begin{figure}
	\centering
	\includegraphics[width=9cm]{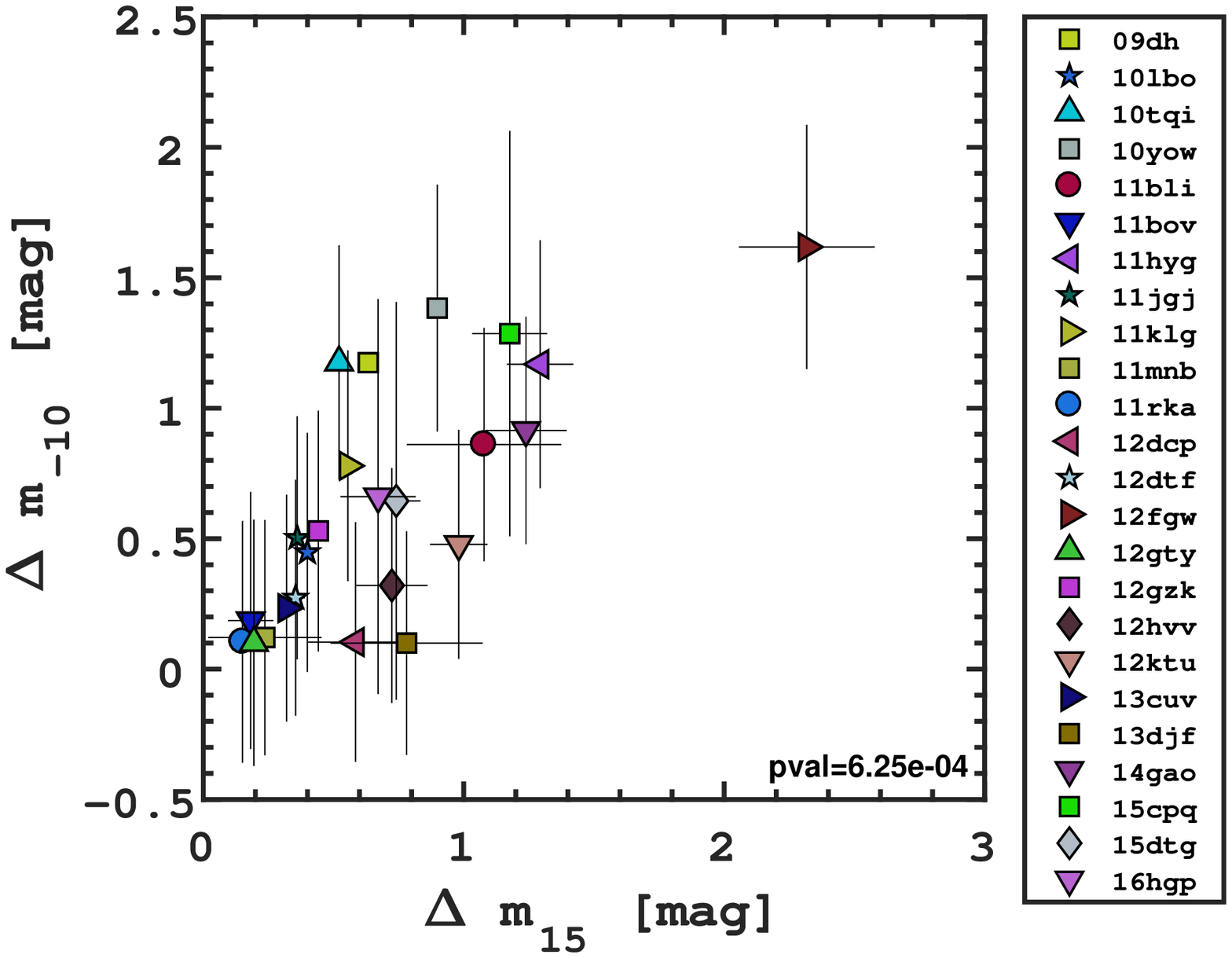}
	\caption{ Bolometric light curve shape:
	$\Delta m_{15}$ vs $\Delta m_{-10}$. The plot does show a correlation, as also found in the r$-$band.
	}
	\label{Fig:bolo_deltas}
\end{figure}

We fit the bolometric light curves with the Contardo function also used in Sect.~\ref{Sect:LCshape}. The best fits are shown in the plot as solid lines in Fig.~\ref{Fig:bolocontardo}.
Following the same analysis as for the $r$ band, this allows us to measure some properties of the shape of the bolometric light curves, such as the peak magnitude, the peak epoch, $\Delta m_{-10}$,  $\Delta m_{15}$ as well as the linear decline slope. We present all these parameters in Table~\ref{Table:LCshape}.
Our sample peak magnitudes span the interval $-16.10$ to $-19.78 \: \mathrm{mag}$ giving an average of $<M_{bol}^{\mathrm{peak}}> = -17.62 \pm 0.94 \: \mathrm{mag}$.
These values are in agreement with the ones available in literature \citep{Drout11,Prentice16,Lyman16,Taddia18b}
We investigated the same correlations as for the $r$ band, they are presented in 
Fig.~\ref{Fig:bolo_deltas}. We find the same correlations as for the r$-$band. We notice that the similarity of relations found among rise and decline parameters in r$-$band and bolometric light curve is likely due to the fact that the flux in r gives a close representation of the bolometric flux at most epochs.
We also estimated the rise times for the bolometric light curves in the same way as we did for the $r$ band and these values are reported in 
Table~\ref{Table:texplo_trise}.

\section{Supernova spectra} \label{Sect:spectra}

This work is focused on the photometric analysis of a sample of SNe Ic, a study of the spectroscopic properties was presented in \citet{Fremling18}. However, one of the aims is to use these light curves to estimate
explosion parameters (Sect.~\ref{Sect:ExplPar}), and in order to break the degeneracy between explosion energy and ejecta mass such an analysis requires
an estimate of the photospheric velocity. This will be
presented here.
We have a total of 177 spectra for the overall sample. All the spectra will become available via the WiseRep archive \citep{Yaron12}. They were already discussed in \citet{Fremling18} and will be released in connection to that paper in a data-release by Fremling et al. (2020, in preparation).

\subsection{Photospheric velocities}\label{Sect:Vel}

\begin{figure*}
	\centering
	\includegraphics[width=18cm]{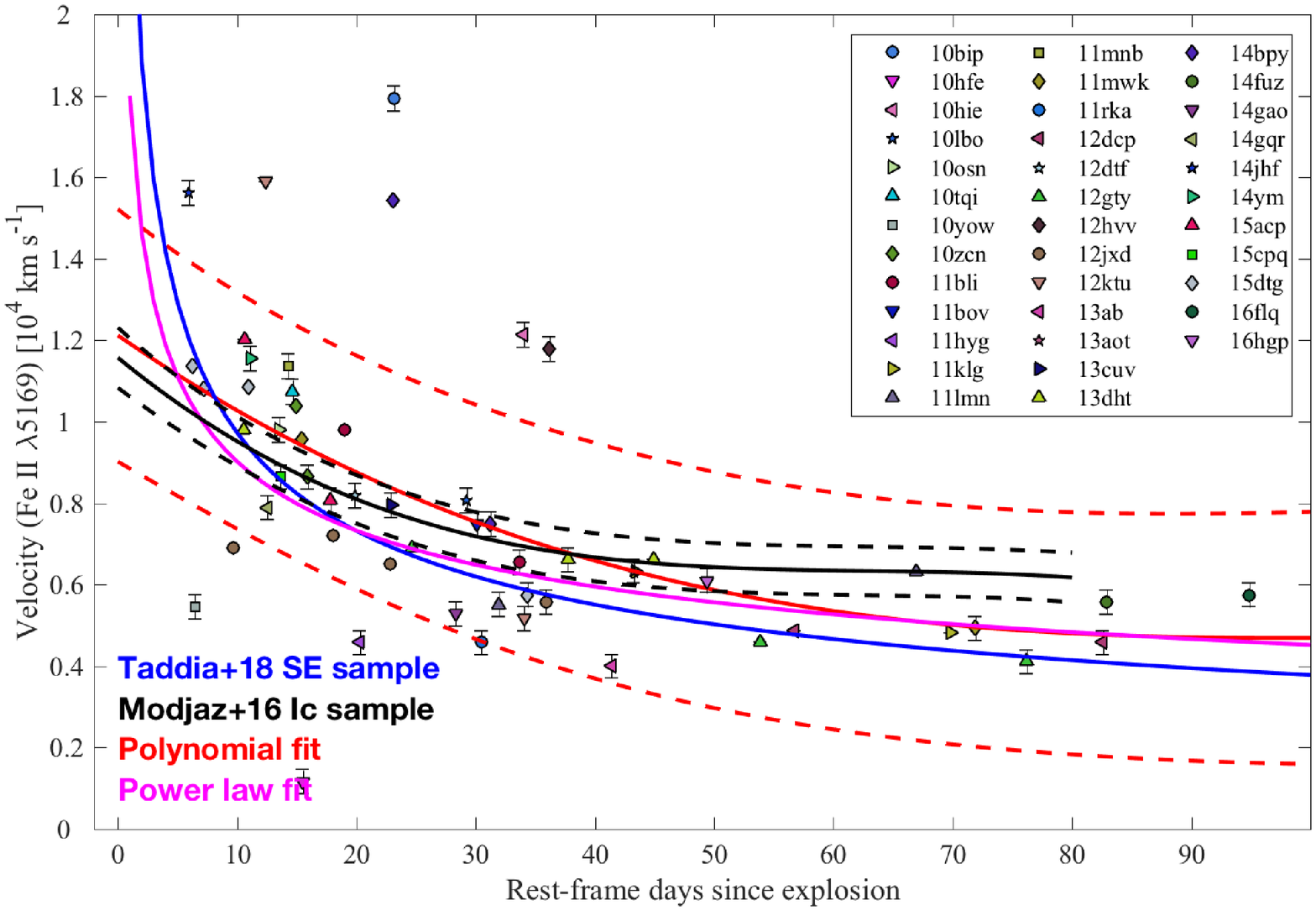}
	\caption{Fe II $\lambda$5169 velocity evolution for 37 SNe~Ic from the sample (see Sect.~\ref{Sect:Vel} for the selection criteria). The magenta solid line represents the power law that fits the evolution with time. The blue solid line represents the trend found by \citet{Taddia18b} and is similar to the one found in this work. The black lines represent the polynomial fit found by \citet{Modjaz16}. As a comparison, we fitted the data also with a polynomial fit, here shown in red.}
	\label{Fig:velFeII}
\end{figure*}

In order to determine the photospheric velocities for the SNe in our sample we estimate the expansion velocities using the \ion{Fe}{II}~$\lambda$5169 line. These velocities were evaluated from the minima of the P-Cygni profile of the \ion{Fe}{II}~$\lambda$5169 for all the available spectra.
For six SNe\footnote{PTF09dh, PTF09ut, PTF11jgj, PTF12cjy, PTF12fgw and iPTF13djf.} 
it was not possible to estimate the \ion{Fe}{II}~$\lambda$5169 velocities as the S/N of the spectra were too low.
Since we aim to build up a functional form for the general trend for normal SNe Ic, we excluded PTF12gzk as it is known from the literature to be a high velocity SN \citep{Horesh13}. We note that this exclusion is only for the purpose to limit the velocity dispersion while determining the functional form, the SN velocities will be estimated and included in the final analysis.
The time evolution of the \ion{Fe}{II}~$\lambda$5169 velocities for the 37 SNe so selected is presented in Fig.~\ref{Fig:velFeII}, where the magenta solid line shows the power law which best represents the trend shown by the overall velocities.

This power-law trend for \ion{Fe}{II}~$\lambda$5169 is in agreement with the trend found by \citet{Taddia18b} for SE SNe, the functional form we found is $v(t) \propto (t-t_{0})^{-0.30}$, where $t_{0}$ represents the explosion epoch.
In Fig.~\ref{Fig:velFeII} we also compare the \ion{Fe}{II}~$\lambda$5169 evolution with the trend found by \citet{Modjaz16}.
Their velocities are lower than our best fits at early epochs.
A polynomial fit is also presented.

The photospheric velocities required to estimate the explosion parameters are the 
\ion{Fe}{II}~$\lambda$5169 velocities at peak. This is not available for every SN in the sample. We then use the general trend found for the overall sample and assume that it represents the velocity evolution for each individual SN. We thus apply our power-law as a template to every SN, shifting it to the available velocity values for the individual SNe. Once it has been shifted, we can extrapolate the value of the \ion{Fe}{II}~$\lambda$5169 velocity at peak.
We estimate the average velocity at peak of the SNe of our sample and adopted this average value to be the velocity for the six SNe for which we were unable to measure the \ion{Fe}{II}~$\lambda$5169 velocity.
The power-law method is also applied to the velocities estimated for PTF12gzk, as these velocities show a similar trend 
but at higher values.
The uncertainties on the peak velocity were assumed to be 10$\%$ of the estimated value.
In this way we get a full set of velocities at peak for the 42 SNe of the sample that will be used in Sect.~\ref{Sect:ExplPar} to estimate the explosion parameters.
The estimated velocities at peak are presented in Table \ref{Table:vel_max}. We notice how PTF10bip presents higher velocities compared to the average of the sample, this could explain the Ic/Ic-BL classification from \citet{Modjaz20}.

 \section{Explosion parameters}\label{Sect:ExplPar}
 
\begin{figure*}
	\centering
	\includegraphics[width=18cm]{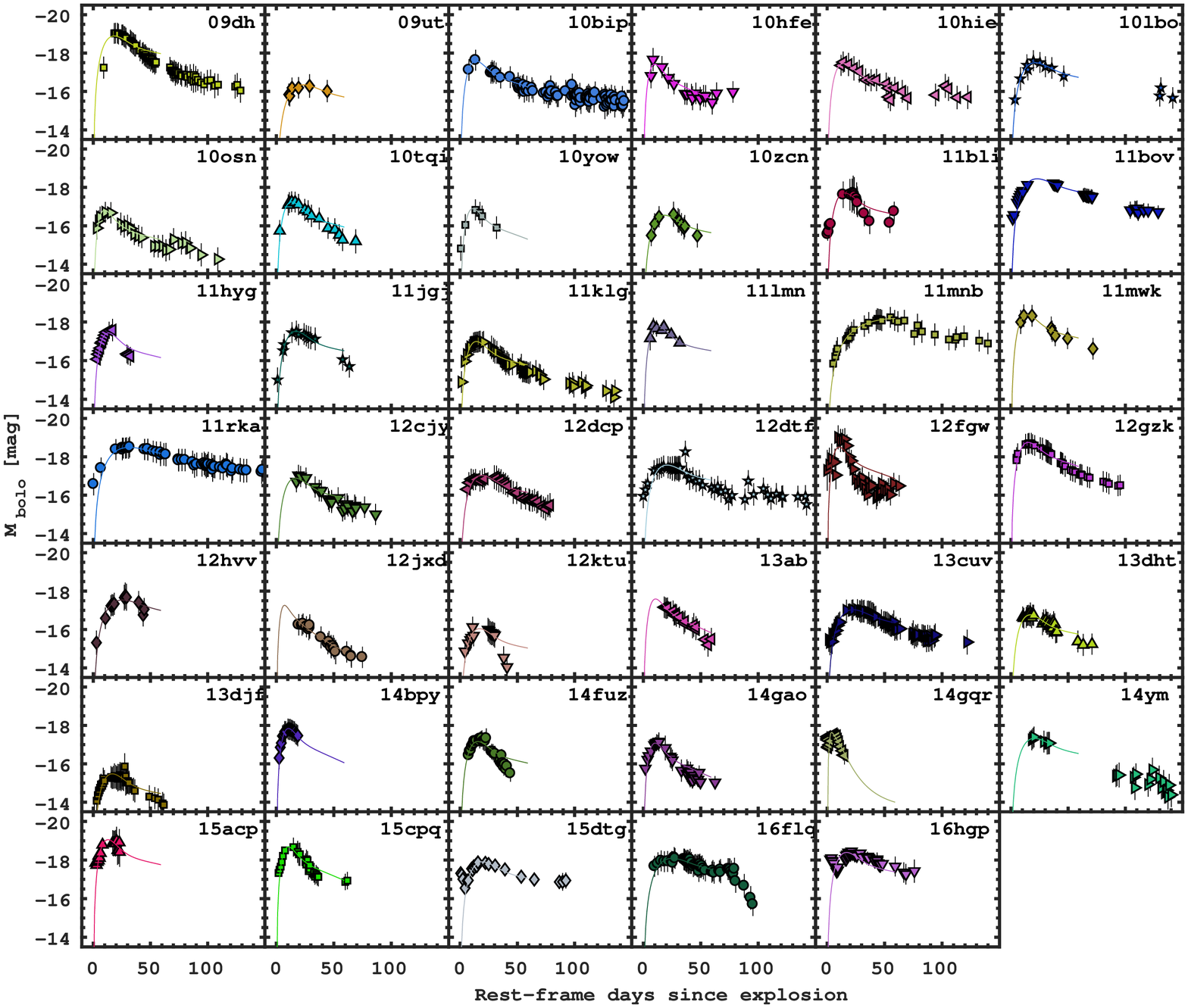}
	\caption{The plot shows the bolometric light curve computed and fitted with Arnett model for the 41 SNe of the sample.}
	\label{Fig:LC_bolo}
\end{figure*}

 In order to estimate the explosion parameters, we fit the bolometric light curves with an Arnett model \citep{Arnett82}. The method we followed to perform the fit is presented in \citet{Taddia18b}.
 We performed the fits on the early epochs of the light curves, $\lesssim$ 60 days after peak, during the photospheric phase of the SNe. The parameters we can estimate from this modelling are the $^{56}$Ni mass ($M_{^{56} \mathrm{Ni}}$), the kinetic energy of the explosion ($E_{K}$) and the ejecta mass ($M_{ej}$). We assume that the SN ejecta have spherical symmetry and uniform density; we also use $E/M = (3/10)V^{2}$, where V is the appropriate ejecta velocity at peak as discussed in Sect.~\ref{Sect:Vel} (see \citet{Valenti08}). We furthermore assumed a constant opacity $\kappa = 0.07 \:  \mathrm{cm^{2} \:  g^{-1}}$, as is often done in the literature for SE SN samples. The Arnett fit for each SN is shown in Fig. \ref{Fig:LC_bolo}.
 
 The estimated values for $M_{^{56}\mathrm{Ni}}$, $M_{ej}$, and $E_{K}$ are listed in Table~\ref{Table:paramexplo}. The uncertainties on $M_{^{56}\mathrm{Ni}}$ are mostly due to the uncertainty in the SN distances. The uncertainties in $M_{ej}$ and $E_{K}$ instead depend mostly on the uncertainty on the expansion velocity.
We notice that the Arnett fit gives a  particular high value of $M_{^{56}\mathrm{Ni}}$ for PTF12gty ($\sim3~\msun$).
We note that PTF12gty has been an outlier for most of the analysis in this work, in particular it has the largest redshift, the highest peak absolute magnitude and longest rise time. It also has the lowest velocity at peak.
We conclude that this SN is most likely a SLSN, as discussed in \cite{DeCia17}
and \cite{Quimby18}.

It will therefore be excluded from the estimates of the average explosion parameters. This reduces the final sample of SNe Ic to 41.
We then obtained average values of $<M_{ej}>=4.50\pm0.79 \; \msun$, $<E_{K}>=1.79 \pm 0.29 \; ~\mathrm{foe}$ (1 foe = 10$^{51}$ erg), and $<M_{^{56}\mathrm{Ni}}>= 0.19 \pm 0.03~\msun$  where the errors are the weighted errors.

\begin{figure*}
	\centering
	\includegraphics[width=18cm]{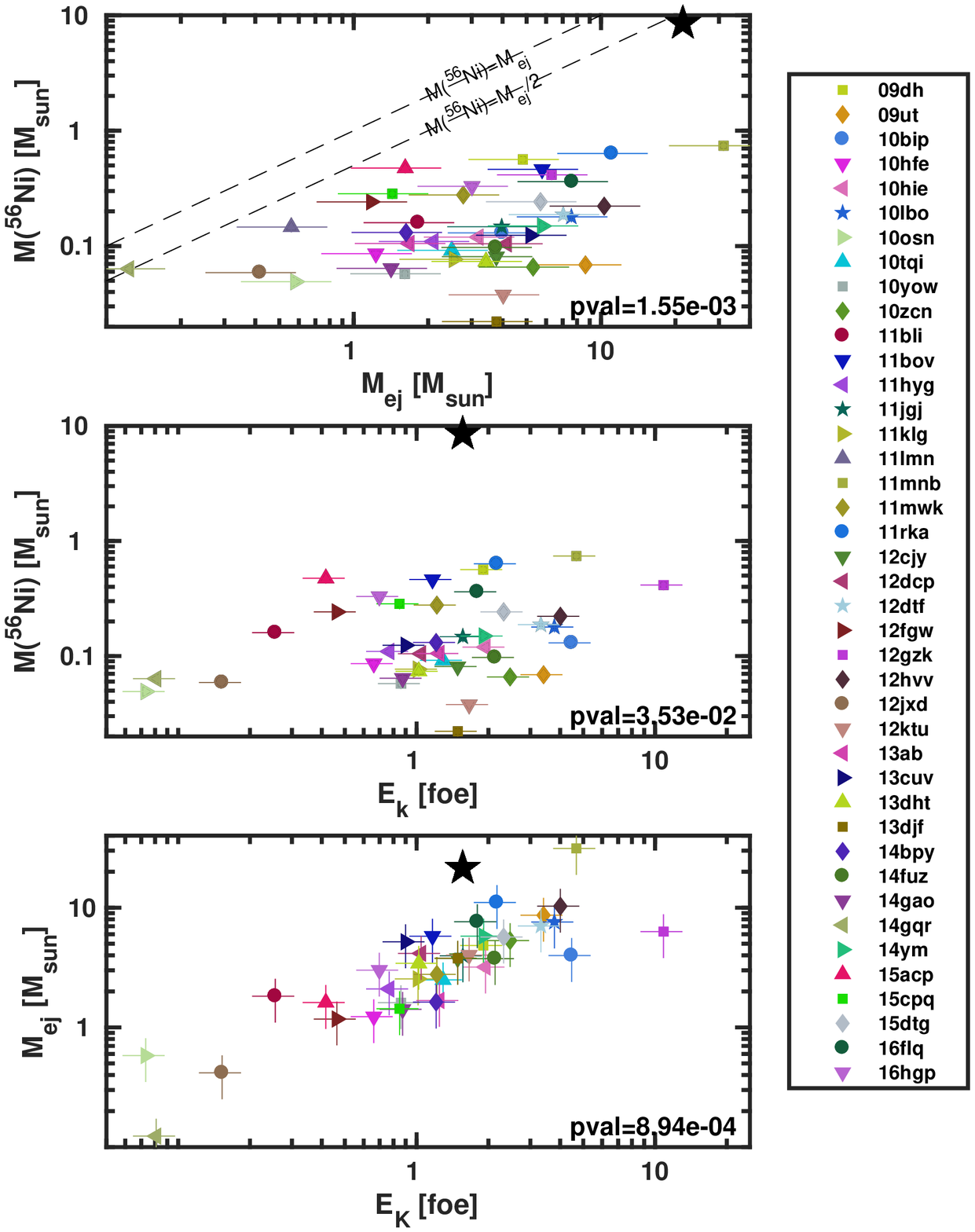}
	\caption{Explosions parameters for 41 SNe Ic plotted against each other. We see clear correlations between the parameters, as quantified by the p-values in the panels. SN PTF12gty has also been represented for completeness with a black star.}
	\label{Fig:explo_param}
\end{figure*}

\begin{figure*}
	\centering
	\includegraphics[width=18cm]{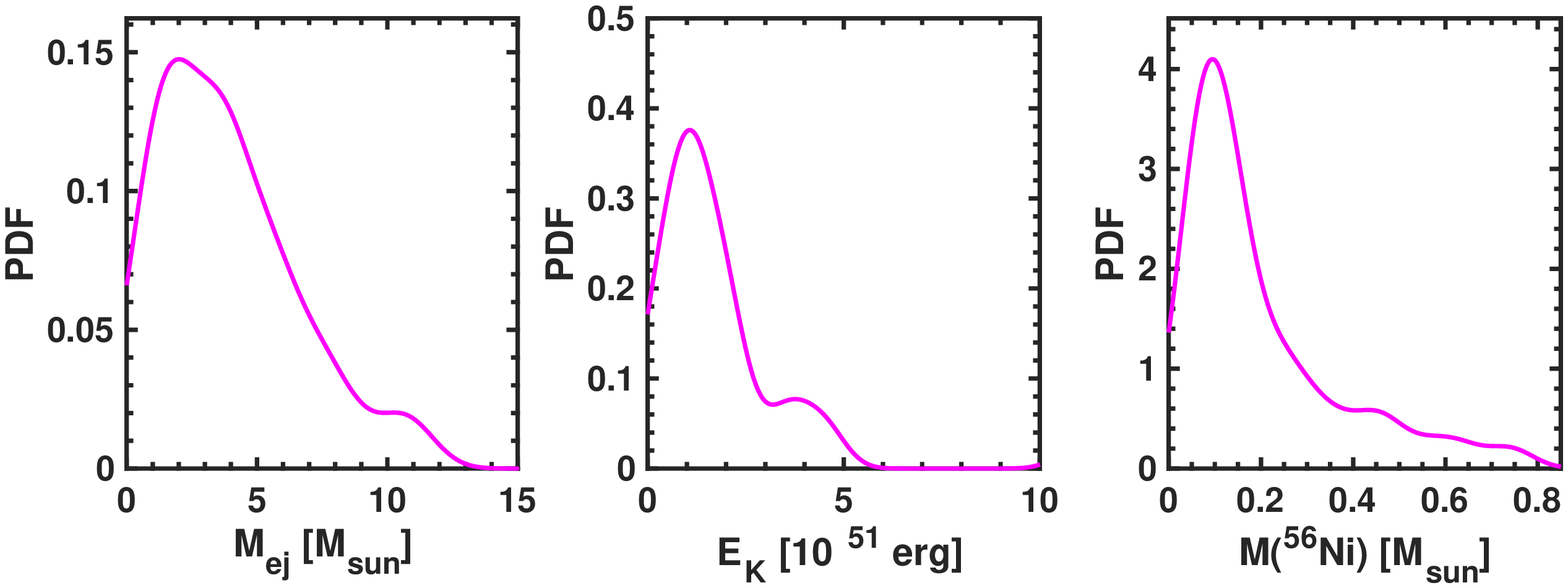}
	\caption{Probability density functions for explosion parameters for our sample of SNe Ic}
	\label{Fig:pdf}
\end{figure*}

In Fig.~\ref{Fig:explo_param} we plot each estimated parameter against the others. We identify a correlation between $M_{ej}$ and $E_{K}$ (see bottom panel). We also notice a correlation between the $M_{ej}$ and $M_{^{56}\mathrm{Ni}}$, and between $M_{^{56}\mathrm{Ni}}$ and $E_{K}$.
e note that the small variation of the velocity's range is possibly driving the correlation between the energy and the ejecta mass.
The probability density function (PDF) of the three explosion parameters are shown in Fig.~\ref{Fig:pdf}.

It shows, for all parameters, that most of the SNe are distributed around a common peak, but there are also evidence for distributions towards higher values in all three parameters.
In Table \ref{Table:SNprop} we present 8 SNe with a type Ibc classification\footnote{PTF09ut, PTF11bli, PTF11lmn, PTF11mwk, iPTF14fuz, iPTF14jhf, iPTF15cpq and iPTF16flq.}. We could not build the bolometric light curve for iPTF14jhf, which leaves us with 7 SNe Ibc. If we exclude these from our sample, we obtained average values of $<M_{ej}>=4.65\pm0.93 \; \msun$, $<E_{K}>=1.88 \pm 0.34 \; ~\mathrm{foe}$, and $<M_{^{56}\mathrm{Ni}}>= 0.19 \pm 0.03~\msun$ where the errors are the weighted errors.
In Sect.~\ref{Sect:sample}, we mentioned that 6 SNe show broader light curves compared to the rest of the sample, which will be discussed separately (Karamehmetoglu et al., in prep). Among these 6 SNe we already excluded PTF12gty as it is most likely a SLSN, leaving us with 5 SNe Ic showing a broad light curve. 
If we exclude these SNe\footnote{PTF11mnb, PTF11rka, iPTF15dtg, iPTF16flq and iPTF16hgp.} we obtain averages $<M_{ej}>=3.57 \pm 0.40 \; \msun$, $<E_{K}>=1.74 \pm 0.33 \; ~\mathrm{foe}$, and $<M_{^{56}\mathrm{Ni}}>= 0.16 \pm 0.02 \; \msun $ where again the errors represent the weighted errors.
Excluding the 5 SNe with broad light curves clearly gives lower average values for
$M_{ej}$ and $M_{^{56}\mathrm{Ni}}$.
If we furthermore exclude from the average the peculiar fast ultrastripped iPTF14gqr \citep{De18} we get $<M_{ej}>=3.67 \pm 0.39 \; \msun$, $<E_{K}>=1.78 \pm 0.32 \; ~\mathrm{foe}$  and $<M_{^{56}\mathrm{Ni}}>= 0.16 \pm 0.02 \; \msun$.
These average values, now based on 34 normal SNe Ic,  are still consistent with the previous ones within the uncertainties.
We estimated the explosion parameters also in the case when the bolometric light curves were built considering the host extinction estimated from the $g-r$ colour evolution. In this case we got $<M_{ej}>=3.17 \pm 0.99$, $<E_{K}>=1.12 \pm 0.23$ , and $<M_{^{56}\mathrm{Ni}}>= 0.77 \pm 0.25$ where again the errors are weighted sigma. The average $<M_{ej}>$, as well as the $<E_{K}>$, are lower but still consistent with the previous measurements within the uncertainties but  $<M_{^{56}\mathrm{Ni}}>$ is higher.
This is not surprising since we got systematically higher values of the extinction with this method and the $M_{^{56}\mathrm{Ni}}$ depends mainly from the peak luminosity.

\section{Discussion and Conclusions} \label{Sect:discussion}

PTF and iPTF allowed for a larger, untargeted, and more homogeneous data set as compared with other SN Ic samples, and for this sample we also have good constraints on the explosion epochs.

We investigated two different methods to estimate the host extinction. First we inspected the spectra to look for Na I D absorption and using \citet{Taubenberger06} we calculated the $E(B-V)$. This method is dependent on the S/N and resolution of the spectrum. The second method is based on the $g-r$ colour evolution and is described in \citet{Stritzinger17}. This method assumes that SE SNe show an intrinsically homogeneous colour evolution in the range $0-20$ days past peak.
We compare the results of these methods in 
Fig.~\ref{Fig:ebmv_vs_ebmv}, which shows that the extinction estimated through the colour evolution is generally higher.

We adopted the extinction estimated from the Na I D for the overall analysis, but also compared how the peak magnitudes would change if we had adapted the other method. The average absolute peak magnitude is $<M_{r}^{\mathrm{peak}}> = -17.71 \pm 0.85 \: \mathrm{mag}$. In case we adopt the extinction from the $g-r$ evolution we get $<M_{r}^{\mathrm{peak}}> = -18.03 \pm 0.79 \: \mathrm{mag}$. 
The effect on the overall peak magnitude distribution are shown in Fig.~\ref{Fig:AbsmDist}, where accounting for higher extinction as suggested by the \citet{Stritzinger17} method shifts the overall sample towards brighter magnitudes.

We investigated the light curve shape in both the $r$ band and for the bolometric light curves. We looked for correlations among the main parameters: magnitude at peak, $\Delta m_{-10}$, $\Delta m_{15}$, $\Delta m_{40}$ and slope. In both cases, we found a correlation between $\Delta m_{15}$ and $\Delta m_{-10}$, implying that slow-rising SNe are also slow decliners. We see a correlation also among $\Delta m_{40}$ vs $\Delta m_{15}$ and among $\Delta m_{40}$ vs $\Delta m_{-10}$.

We fitted the bolometric light curves with an Arnett model \citep{Arnett82} to estimate the explosion parameters. We  obtained average values of $<M_{ej}>=4.39\pm0.31 \; \msun$, $<E_{K}>=1.71 \pm 0.16 \; ~\mathrm{foe}$, and $<M_{^{56}\mathrm{Ni}}>= 0.19 \pm 0.05~\msun$, when including all the 41
SNe Ic for which we could estimate these parameters.
We searched for correlations among the explosion parameters and  identify a correlation between  $M_{ej}$ and $E_{K}$. 
We also notice a correlation between the $M_{ej}$ and $M_{^{56}\mathrm{Ni}}$, and between $M_{^{56}\mathrm{Ni}}$ and $E_{K}$.

\subsection{Comparison with the literature}

Some of the SNe in this sample have already been discussed in the literature, and we will here compare our results with those available in  these publications.
SNe PTF09dh, PTF11bli, PTF11jgj, PTF11klg, PTF11rka and PTF12gzk were presented in \citet{Prentice16}. Our estimated $M_{^{56}\mathrm{Ni}}$ values 
for these SNe are in agreement with the ones provided in their work.
PTF12gzk is discussed in \citet{BenAmi12}, in which they noted that this SN showed some aspects 
in-between SNe Ic and SNe Ic-BL. They conclude that the mass of the progenitor star is $25-35 \: \msun$.
We get quite high values for the ejecta mass which might point towards a massive progenitor star as found by \citet{BenAmi12}.
PTF11bov is also known as SN 2011bm and was presented in \citet{Valenti12} where they infer an initial mass for the progenitor of $30-50 \: \msun$. The ejecta mass we derive is close to the lower end of the interval they present in their work.
iPTF15dtg was first introduced in \citet{Taddia16} and investigated further in \citet{Taddia19a}. In their first work they concluded that the peculiar long rise of this SN was most likely due to an extended envelope around the progenitor star, which they claim was a massive ($> 35~\msun$) Wolf-Rayet star. The overall explosion parameters we estimated using the Arnett model are somewhat consistent with their lower values. In the subsequent paper, they accounted for additional peculiar behaviour of the SN at late times, which was explained by a combination of radioactive and magnetar powering which leads to a lower estimate of $M_{ej}$ when compared with our estimate. \\
iPTF14gqr was presented in \citet{De18} where they concluded that the best interpretation for this fast event is an ultra-stripped SN. 
We also obtained low values for the ejecta mass and kinetic energy, in agreement with the scenario presented in \citet{De18}.
iPTF11mnb was presented in a separate paper as a SN Ic from a massive progenitor ($85 \: \msun$; \citealt{Taddia18a}).
Our estimates also show high values for the explosion parameters pointing towards a massive progenitor star.
iPTF12gty was classified as a SLSN by \citet{Quimby18} and further investigated by \citet{DeCia17}. Our spectral classification was pointing towards a SN Ic classification but this SN is a clear outlier in the sample in many ways. In particular, when applying the Arnett fit to estimate the explosion parameters we get a very high value of the $^{56}\mathrm{Ni}$ mass. We therefore conclude that iPTF12gty is most likely a super-luminous SN.

Our $r$-band absolute magnitudes span the interval $-15.54$ to $-19.81 \: \mathrm{mag}$, with an average of $<M_{r}> = -17.71 \pm 0.85 \: \mathrm{mag}$.
The ranges available in the literature are $M_{r}^{\mathrm{peak}} = -18.26 \pm 0.21 \: \mathrm{mag}$  \citep{Taddia15}; $M_{r}^{\mathrm{peak}} = -17.64 \pm 0.26 \: \mathrm{mag}$  \citep{Taddia18b} and  $M_{R}^{\mathrm{peak}} = -18.3 \pm 0.6 \: \mathrm{mag}$ \citep{Drout11}.
The average peak magnitude in the $r$ band estimated for our sample is in agreement with the ones from literature.
We compared our values also with the (i)PTF sample of SNe Ic-BL \citep{Taddia19b} where the peak magnitudes show a brighter  average of $-18.7 \pm 0.7 \: $mag.

We analysed the shapes of the $r$-band light curves and of the bolometric light curves, searching for correlations among the different parameters. We identified a correlation in the $r$ band between $\Delta m_{15}$ and $\Delta m_{-10}$ which is in agreement with  the results from\citet{Taddia19b,Drout11}. 
We note that the fact that the fast risers are also the fast decliners is not trivially true. There could well be different physical circumstances determining the rise and the decline from peak, for example the mixing out of radioactive nickel will affect the steepness of the rising light curve whereas the time scale for the decline may be more determined by the ejecta mass and composition.
We did not find any Phillips-like relation and this is in agreement with previous works \citep{Taddia19b,Lyman16,Drout11}.

\begin{figure*}
	\centering
	\includegraphics[width=18cm]{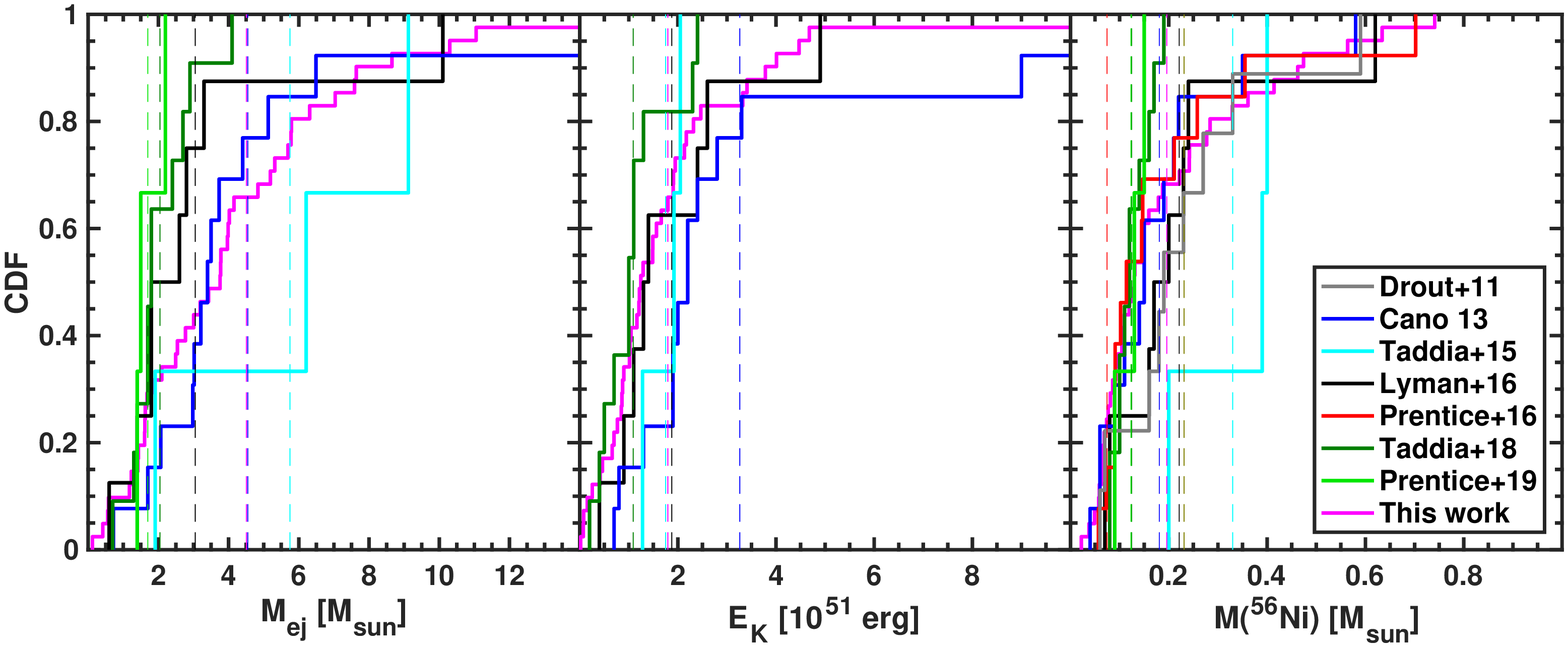}
	\caption{Cumulative distribution functions for the explosion parameters compared to those of other samples in the literature. Dashed lines represent the average values for each sample from literature.
	}
	\label{Fig:cdf}
\end{figure*}

We compared the estimated average values for the explosion parameters of the 41 (i)PTF SNe Ic with the ones available in the literature.
\citet{Drout11} presented $M_{^{56}\mathrm{Ni}}$ values for 9 SNe Ic.
In \citet{Cano13} the explosion parameters for 13 SNe Ic are presented.
\citet{Taddia15} analysed three events, while the \citet{Lyman16} sample contains 8 SNe Ic. A total number of 13 SNe Ic was presented in \citet{Prentice16}. \citet{Taddia18b}  presented 11 SNe Ic and in \citet{Prentice19} three SNe Ic are included.
Our (i)PTF sample with 41 SNe Ic
therefore by far represents the largest sample of SNe Ic available where the explosion parameters have been estimated.
We estimated the cumulative distribution functions (CDF) of the explosions parameters, 
and compared it to the available studies in the literature. 
The results of these comparisons are shown in Fig.~\ref{Fig:cdf}.

We also report the average values and their
standard deviations for the estimated explosion parameters from the different samples in 
Table~\ref{Table:paramexplocomp}.
We also compared the $M_{^{56}\mathrm{Ni}}$ estimated by \citet{Meza20} for 6 SNe Ic using Arnett model, and they get an average value lower than the one found in this work.

We searched for correlations among the explosions parameters (see Fig.~\ref{Fig:explo_param}), and
identify a correlation between  $M_{ej}$ and $E_{K}$.
We also notice correlations between  $M_{ej}$ and $M_{^{56}\mathrm{Ni}}$, and between $M_{^{56}\mathrm{Ni}}$ and $E_{K}$.
These correlations were also observed in other SE SN studies \citep{Taddia19a,Taddia18a,Lyman16}.
The strong correlation between ejecta mass and kinetic energy is, just as noted by \cite{Lyman16}, in fact mainly driven by the ejecta mass. The range in ejecta mass is much larger than the variation in velocity, and this is driving the relation. There is no correlation between ejecta mass and photosperic velocity.

\subsection{Implications for progenitors}
The PDF of the different explosion parameters are shown in Fig.~\ref{Fig:pdf}. 
The $E_{K}$ shows a first strong peak for energies lower than $3 \: \mathrm{foe}$ and the $M_{^{56}\mathrm{Ni}}$ distribution shows a clear peak at values lower than $0.3 \: \msun$.
The PDF of the $M_{ej}$ shows a first peak for values lower than $5 \: \msun$ and shows indication for additional peak(s) towards higher mass.
A similar analysis for $M_{ej}$ was presented in \citet{Lyman16} and from a comparison with expectations from models they concluded that 
since the peak of ejecta masses is rather low, this indicates that
the majority of SE SNe originate from not too massive stars, assuming the remnant is a neutron star. Moreover, since such stars are unable to get rid of all of their outer hydrogen and helium layers solely from mass-loss from winds, they must likely have been born and stripped in a binary system. 
The trend we see here for our larger sample would lead to a similar conclusion. We notice that the PDF presented in \citet{Lyman16} did not show a very pronounced secondary peak. This is due to the presence of SNe with broad light curves in our sample, which could  arise from more massive star progenitors. 
We note that the apparent secondary peak in the PDF at $\sim 10~\msun$ of ejected mass seems to be compatible with single, massive WR progenitors. These SNe will be discussed into more detail by Karamemehtoglu et al., (in prep.).
\\

We have presented the sample of Type Ic supernovae collected by (i)PTF over a period of $\sim7$ years. The final sample of SNe that also have pre$-$peak photometry is made up of 44 objects, which we analyse in terms of light-curve and explosion parameters. This is the largest such sample to date. Our main results confirm trends seen in previous articles based on smaller and less homogeneous samples. Although our data are not always fantastic for individual SNe, the bulk sample provides a good picture of the overall properties of this class of extremely stripped supernovae.

The moderate ejecta masses remain a challenge for scenarios involving single very massive stars, as already proposed by \citet{Lyman16}, and corroborate discussions on the need for binary star evolution to produce most of the Type Ic SNe. Indications for a population of more massive progenitors are also seen. The ejected masses of radioactive nickel are 
$\sim0.2 \msun$, which is more than current neutrino driven explosion models \citep{Ertl20} can easily accomplish. As mentioned, the correlation between ejecta mass and energy is largely spurious - but the correlation between ejecta mass and mass of radioactive nickel appears to be robust. It is statistically significant even if we exclude the most massive object that drives the correlation, and is something that a generic explosion model would have to explain.

There is hope for better understanding of these explosions from the
observational perspective. The Zwicky Transient Facility \citep{Bellm19}
that has taken over on the P48 telescope after (i)PTF enable superior light curves also of Type Ic SNe. Over the first years, this survey has already observed almost 100 Type Ic SNe, and a fair fraction of these have better sampled LC:s than the sample we have presented here. We look forward to analysing these new data.

\section{Acknowledgements}

The Oskar Klein Centre was funded by the Swedish Research
Council. C.B. gratefully acknowledge support from the
Knut and Alice Wallenberg Foundation, and from the Wennergren Foundation (PI JS).

The intermediate Palomar Transient Factory project is a scientific collaboration among the California Institute of Technology, Los Alamos National Laboratory, the University of Wisconsin, Milwaukee, the Oskar Klein Center, the Weizmann Institute of Science, the TANGO Program of the University System of Taiwan, and the Kavli Institute for the Physics and Mathematics of the Universe. LANL participation in iPTF is supported by the US Department of Energy as a part of the Laboratory Directed Research and Development program.

The data presented herein were obtained in part with ALFOSC, which is provided by the Instituto de Astrofisica de Andalucia (IAA) under a joint agreement with the University of Copenhagen and NOTSA. 

IA is a CIFAR Azrieli Global Scholar in the Gravity and the Extreme Universe Program and acknowledges support from that program, from the Israel Science Foundation (grant numbers 2108/18 and 2752/19), from the United States - Israel Binational Science Foundation (BSF), and from the Israeli Council for Higher Education Alon Fellowship.

AGY’s research is supported by the EU via ERC grant No. 725161, the ISF GW excellence center, an IMOS space infrastructure grant and BSF/Transformative and GIF grants, as well as The Benoziyo Endowment Fund for the Advancement of Science, the Deloro Institute for Advanced Research in Space and Optics, The Veronika A. Rabl Physics Discretionary Fund, Paul and Tina Gardner, Yeda-Sela and the WIS-CIT joint research grant;  AGY is the recipient of the Helen and Martin Kimmel Award for Innovative Investigation

\bibliographystyle{aa}

\clearpage

\begin{landscape}
\begin{table}
	\caption{List of the 44 SNe Ic in the sample used in this work with their coordinates, redshifts, distances and extinction estimates.
	\label{Table:SNprop}}
	\begin{footnotesize}
		\begin{tabular}{lcccccccccc}
			\hline
			SN  & Type & RA & Dec & z & Distance modulus & Distance & $E(B-V)_{MW}$ & $E(B-V)_{host}$ &$E(B-V)_{host}$ &$E(B-V)_{host}$\\
			& & &  & &  &  &  & from Na I D$^{a}$& from Na I D$^{b}$ & from $g-r$ \\
			& & (h m s)  & ($^{\circ}$ ' ") & & (mag) & (Mpc) & (mag) & (mag) & (mag) & (mag)\\
			\hline
		09dh   & Ic		& 14:44:42.07	&	+49:43:44.9  &  0.07$^{1}$   &	37.48 	&  313.91   &  0.0213	&	  0.16     & 0.21 (0.07) &  0.14(0.14)  \\
		09ut   & Ibc	& 14:12:07.64	&	+74:45:46.0  &  0.04$^{1}$   &	36.33 	&  184.50   &  0.0260	&	  0.00     & 0.01 (0.01) &  0.24(0.25)  \\
		10bip  & Ic		& 12:34:10.52	&	+08:21:48.5  &  0.051        &	36.77 	&  225.55   &  0.0159	&	  0.06     & 0.04 (0.01) &  0.00(0.00)  \\
		10hfe  & Ic		& 12:32:05.16	&	+66:24:23.9  &  0.049        &	36.68 	&  216.38   &  0.0157	&	  0.00     & 0.01 (0.01) &  0.00(0.00)  \\
		10hie  & Ic		& 17:02:27.71	&	+28:31:57.6  &  0.067$^{2}$  &	37.38 	&  299.81	&  0.0464	&	  0.02     & 0.02 (0.01) &  0.33(0.25)  \\
		10lbo  & Ic	    & 12:59:14.79   &   +61:27:02.2  &  0.053$^{2}$  &  36.85 	&  234.74   &  0.0138	&     0.00     & 0.01 (0.01) &  0.00(0.00)  \\     
		10osn  & Ic		& 23:23:07.98	&	+17:30:29.1  &  0.038        &	36.11 	&  166.43   &  0.0361	&	  0.00     & 0.01 (0.01) &  0.06(0.12) \\
		10tqi  & Ic		& 23:20:47.73	&	+18:54:17.3  &  0.038        &	36.14 	&  169.13   &  0.0347	&	  0.16     & 0.21 (0.07) &  0.18(0.25) \\
		10yow  & Ic		& 21:54:23.30	&	+15:09:20.7  &  0.024        &	35.13 	&  106.34   &  0.0902	&	  0.18     & 0.27 (0.10) &  0.22(0.25) \\
		10zcn  & Ic		& 23:19:14.39	&	+26:03:11.6  &  0.02$^{1}$   &	34.78 	&  90.25    &  0.0794	&	  0.12     & 0.10 (0.04) &  0.64(0.25) \\
		11bli  & Ibc	& 14:02:16.18	&	+33:39:41.5  &  0.034        &	35.88 	&  149.80   &  0.0124	&	  0.10     & 0.07 (0.03) &  0.49(0.16) \\
		11bov  & Ic		& 12:56:53.94	&	+22:22:28.1  &  0.021        &	34.90 	&  95.62    &  0.0289	&	  0.05     & 0.03 (0.01) &  0.00(0.00) \\
		11hyg  & Ic		& 23:27:57.34	&	+08:46:38.0  &  0.029        &	35.50 	&  125.81   &  0.0509	&	  0.10     & 0.08 (0.03) &  0.26(0.14) \\
		11jgj  & Ic		& 16:31:32.35	&	+62:06:09.5  &  0.040$^{2}$  &	36.22 	&  175.45   &  0.0257	&	  0.22     & 0.61 (0.22) &  0.26(0.12) \\
		11klg  & Ic		& 22:07:09.92	&	+06:29:08.7  &  0.026        &	35.31 	&  115.15   &  0.0754	&	  0.00     & 0.01 (0.01) &  0.28(0.18) \\
		11lmn  & Ibc	& 17:30:16.33	&	+26:59:34.0  &  0.090$^{2}$  &	38.07 	&  411.33   &  0.0439	&	  0.00     & 0.01 (0.01) &  0.00(0.00) \\
		11mnb  & Ic		& 00:34:13.25	&	+02:48:31.4  &  0.060$^{2}$  &	37.14 	&  268.51   &  0.0157	&	  0.08     & 0.05 (0.02) &  0.08(0.17) \\
		11mwk  & Ibc	& 21:35:01.39	&	+00:07:16.0  &  0.121$^{2}$  &	38.76 	&  564.78   &  0.0457	&	  0.02     & 0.02 (0.01) &  0.00(0.00) \\    
		11rka  & Ic		& 12:40:44.87	&	+12:53:21.4  &  0.074$^{2}$  &	37.62 	&  334.70   &  0.0295	&	  0.08     & 0.05 (0.02) &  0.29(0.13) \\    
		12dtf  & Ic		& 17:14:43.72	&	+45:18:19.0  &  0.061$^{2}$  &	36.43 	&  193.58   &  0.0347	&	  0.00     & 0.01 (0.01) &  0.36(0.16) \\    
		12dcp  & Ic		& 16:12:56.12	&	+32:30:43.2  &  0.031        &	35.65 	&  134.74   &  0.0204	&	  0.00     & 0.01 (0.01) &  0.38(0.17) \\     
		12cjy  & Ic		& 14:34:27.31	&	+51:49:03.7  &  0.044        &	37.17 	&  271.77   &  0.0082	&	  0.04     & 0.03 (0.01) &  0.00(0.00) \\
		12fgw  & Ic		& 15:53:27.12	&	+33:21:04.2  &  0.055        &	36.94 	&  243.96   &  0.0287	&	  0.00     & 0.01 (0.01) &  0.00(0.00) \\     
		12gty  & Ic		& 16:01:15.23	&	+21:23:17.4  &  0.17$^{1}$   &	39.64 	&  847.56   &  0.0589	&	  0.03     & 0.02 (0.01) &  0.04(0.16) \\
		12gzk  & Ic		& 22:12:41.53	&	+00:30:43.1  &  0.013$^{2}$  &	33.86 	&  59.20    &  0.0432	&	  0.14     & 0.14 (0.05) &  0.21(0.12) \\
		12hvv  & Ic		& 21:45:46.45	&	-00:03:25.1  &  0.029$^{2}$  &	35.50 	&  126.14   &  0.0715	&	  0.08     & 0.05 (0.02) &  0.70(0.22) \\
		12jxd  & Ic		& 09:37:29.82	&	+23:09:50.2  &  0.025        &	35.21 	&  110.18   &  0.0258	&	  0.32     & 0.30 (1.14) &  0.43(0.25) \\
		12ktu  & Ic		& 04:26:20.58	&	-10:06:12.5	 &  0.031        &	35.65 	&  135.05   &  0.0567	&	  0.08     & 0.05 (0.02) &  0.34(0.12) \\ 
		13ab   & Ic		& 12:38:12.51	&	+07:09:02.0  &  0.048        &	36.63 	&  211.81   &  0.0183	&	  0.04     & 0.03 (0.01) &  0.00(0.00) \\
		13aot  & Ic		& 13:18:26.09	&	+31:28:09.9  &  0.018        &	34.53 	&  80.65    &  0.0103	&	  0.04     & 0.02 (0.01) &   --  \\
		13cuv  & Ic		& 01:53:20.32	&	+35:50:19.3  &  0.049$^{2}$  &	36.48 	&  198.13   &  0.0605	&	  0.02     & 0.04 (0.01) &  0.26(0.13) \\
		13dht  & Ic		& 23:44:58.88	&	+09:55:24.8  &  0.040$^{2}$  &	36.22 	&  175.45   &  0.0658	&	  0.06     & 0.02 (0.01) &  0.20(0.17) \\
		13djf  & Ic		& 23:33:38.73	&	+08:48:44.6  &  0.021        &	34.70 	&  87.07    &  0.0690	&	  0.03     & 0.01 (0.01) &  0.57(0.25) \\
		14bpy  & Ic		& 15:33:08.66	&	+17:17:53.3  &  0.04$^{1}$   &	36.48 	&  198.13   &  0.0361	&	  0.00     & 0.03 (0.01) &  0.00(0.00) \\     
		14gao  & Ic		& 00:57:40.21	&	+43:47:35.2  &  0.018        &	36.43 	&  193.58   &  0.0688	&	  0.04     & 0.00 (0.01) &  0.55(0.14) \\
		14gqr  & Ic		& 23:33:27.95	&	+33:38:46.1	 &  0.063$^{2}$  &	34.45 	&  77.64 	&  0.0800	&	  0.41     & 0.01 (0.01) &  0.00(0.00) \\
		14fuz  & Ibc	& 01:05:30.18	&	+02:51:42.0  &  0.044        &	37.24 	&  281.09   &  0.0235	&	  0.00     & 0.00 (0.01) &  0.00(0.00) \\
		14jhf  & Ibc	& 08:00:33.06	&	+18:15:35.9  &  0.053$^{2}$  &	36.90 	&  239.35   &  0.0202	&	  0.31     & 0.17 (0.98) &   --  \\
		14ym   & Ic		& 17:46:40.45	&	+58:38:11.1  &  0.031        &	35.68 	&  136.83   &  0.0400	&	  0.15     & 0.12 (0.06) &  0.22(0.16) \\
		15acp  & Ic		& 14:53:56.34	&	+59:02:51.4  &  0.138$^{2}$  &	39.06 	&  648.56   &  0.0094	&	  0.13     & 0.01 (0.04) &  0.33(0.14) \\
		15cpq  & Ibc	& 00:13:43.16	&	+00:09:43.2  &  0.066$^{2}$  &	37.35 	&  295.12   &  0.0238	&	  0.10     & 0.07 (0.03) &  0.16(0.14) \\
		15dtg  & Ic		& 02:30:20.05	&	+37:14:06.7  &  0.052$^{2}$  &	36.83 	&  231.98   &  0.0549	&	  0.00     & 0.01 (0.01) &  0.11(0.12) \\
		16flq  & Ibc	& 00:28:36.54	&	-01:33:03.3  &  0.06$^{1}$   &	37.13 	&  266.65   &  0.0190	&	  0.24     & 0.74 (0.27) &  0.23(0.12) \\
		16hgp  & Ic		& 00:12:06.41	&	+32:11:50.9  &  0.079$^{2}$  &	37.76 	&  356.57   &  0.038	&	  0.00     & 0.01 (0.01) &  0.06(0.14) \\
			\hline
			
\end{tabular} \\
$^{a}$ $E(B-V)$ estimate with Na I D using the \citet{Taubenberger06} relation. Adopted uncertainties of $\Delta E(B-V) = \: 0.2 \: \mathrm{mag}$ \\
$^{b}$ $E(B-V)$ estimate with Na I D using the \citet{Poznanski12} relation \\
$^{1}$Redshift estimated from SNID \\
$^{2}$Redshift estimated from narrow host line
\\[1.5ex]
\end{footnotesize}
\end{table}
\end{landscape}

\begin{table*}
	\caption{Parameters for the analysis of the lightcurve shape in the $r$ band for the 40 SNe, uncertainties are in parentesis.} \label{Table:LCshape}
	\begin{footnotesize}
		\begin{tabular}{lccccc}
			\hline
			SN & $t_{r}^{max}$ & $m_{r}^{peak}$  & $\Delta m_{15} (r)$ & $\Delta m_{-10} (r)$ & $\Delta m_{40} (r)$ \\
			& (JD) & (mag) & (mag) & (mag) & (mag) \\
			\hline
  09dh & 2454954.43(0.11) & 18.65(0.01) & 0.62(0.02) & 0.71(0.08)  &  1.70(0.04)  \\
  09ut & 2455014.29(1.97) & 19.91(0.11) & 0.20(0.05) &  ---        &  ---  \\
  10bip & 2455231.72(0.12) & 19.38(0.05) & 0.43(0.01) &  ---       &  1.17(0.04)  \\
  10hfe & 2455341.19(1.41) & 20.00(0.11) & 0.50(0.03) &  ---       &  0.94(0.11)  \\
  10hie & 2455336.27(0.01) & 19.71(0.08) & 0.73(0.05) &  ---       &  1.62(0.13)  \\
  10lbo & 2455377.10(0.50) & 19.37(0.07) & 0.29(0.01) & 0.26(0.01) &  0.81(0.01)  \\
  10osn & 2455398.18(0.70) & 19.27(0.07) & 0.65(0.01) &  ---       &  1.51(0.08)  \\
  10tqi & 2455443.58(0.30) & 19.05(0.08) & 0.55(0.02) & 0.80(0.04) &  1.64(0.14)  \\
  10yow & 2455480.75(0.88) & 18.61(0.12) & 0.84(0.05) & 1.29(0.17) &  ---  \\
  10zcn & 2455487.63(0.21) & 18.42(0.05) & 0.65(0.00) &  ---       &  ---  \\
  11bli & 2455646.01(1.04) & 18.61(0.05) & 0.49(0.03) &  ---       &  1.18(0.05)  \\
  11bov & 2455685.29(0.16) & 16.85(0.03) & 0.19(0.01) & 0.07(0.00) &  0.70(0.05)  \\
  11hyg & 2455752.40(0.15)$^{a}$ & --- & --- & --- & --- \\
  11jgj & 2455786.34(0.88) & 19.03(0.05) & 0.37(0.08) & 0.34(0.04) &  1.52(0.16)  \\
  11klg & 2455809.06(0.03) & 18.25(0.01) & 0.76(0.01) & 0.62(0.02) &  1.66(0.04)  \\
  11lmn & 2455813.44(1.31)$^{a}$ & --- & --- & --- & --- \\
  11mnb & 2455861.35(1.17) & 18.86(0.02) & 0.37(0.13) & 0.19(0.07) &   0.81(0.05)  \\
  11mwk & 2455829.79(0.93) & 20.48(0.19) & 0.48(0.05) &  ---       &   1.24(0.04)  \\
  11rka & 2455930.57(0.43) & 19.11(0.03) & 0.14(0.02) & 0.11(0.01) &   0 .60(0.05)  \\
  12cjy & 2456026.13(0.90) & 19.59(0.06) & 0.53(0.04) &  ---       &   1.31(0.12)  \\
  12dcp & 2456048.16(1.59) & 18.74(0.07) & 0.32(0.06) & 0.07(0.03) &  1.22(0.05)  \\
  12dtf & 2456056.13(0.49) & 19.64(0.05) & 0.41(0.02) & 0.38(0.03) &  1.12(0.02)  \\
  12fgw & 2456093.36(0.06) & 19.30(0.08) & 1.52(0.05) & 1.45(0.05) &  1.82(0.05)  \\
  12gty & 2456171.93(0.80) & 20.08(0.05) & 0.41(0.06) & 0.11(0.02) &  1.11(0.14)  \\
  12gzk & 2456152.44(0.03) & 15.37(0.01) & 0.46(0.03) & 0.42(0.01) &  1.49(0.01)  \\
  12hvv & 2456165.97(0.12) & 18.34(0.02) & 0.14(0.01) & 1.21(0.07) &  --- \\
  12jxd & 2456229.95(0.31) & 19.33(0.07) & 0.67(0.03) &  ---       &  1.54(0.04)  \\
  12ktu & 2456245.37(0.40) & 19.37(0.03) & 1.06(0.14) & 0.89(0.13) &  ---  \\
  13ab & 2456333.82(0.87) & 19.85(0.06) & 0.39(0.02) &  ---        &  ---  \\
  13aot & 2456416.55(0.35) & 17.88(0.04) & 0.43(0.01) & 0.16(0.01) &  1.11(0.01)  \\
  13cuv & 2456537.31(0.29) & 19.34(0.07) & 0.43(0.04) & 0.32(0.01) &  1.13(0.06)  \\
  13dht & 2456551.36(0.89) & 19.49(0.07) & 0.51(0.04) &  ---       &  1.41(0.11)  \\
  13djf & 2456557.86(0.61) & 19.40(0.09) & 0.60(0.04) & 0.65(0.14) &  1.46(0.09)  \\
  14bpy & 2456835.60(0.68) & 18.89(0.08) &   ---    & 1.05(0.14)   &  ---  \\
  14fuz & 2456927.38(0.05) & 19.05(0.04) & 0.43(0.05) &  ---       &  ---  \\
  14gao & 2456925.59(0.83)$^{a}$ & --- & --- & --- & --- \\
  14gqr & 2456951.88(0.05) & 20.02(0.04) &   ---     &  ---        &  ---  \\
  14jhf & 2457034.93(0.12) & 19.14(0.15) &   ---    & 0.16(0.02)   &  ---  \\
  14ym & 2456715.42(1.10) & 18.49(0.10) & 0.38(0.05) &  ---        &   1.07(0.03)  \\
  15acp & 2457122.24(0.26) & 20.33(0.08) & 0.75(0.04) & 0.64(0.03) &  ---  \\
  15cpq & 2457283.61(0.40)$^{a}$ & --- & --- & --- & --- \\
  15dtg & 2457372.83(0.06) & 18.93(0.01) & 0.09(0.02) & 0.05(0.02) &  0.47(0.01)  \\
  16flq & 2457644.56(0.30) & 19.24(0.07) & 0.16(0.01) & 0.17(0.01) &  0.52(0.01)  \\
  16hgp & 2457712.03(0.07) & 19.36(0.04) & 0.25(0.01) & 0.16(0.05) &  0.66(0.07)   \\
			\hline
		\end{tabular} \\
		$^{a}$ Epoch of $r-$band peak estimated from the $g-$band peak, considering the average shift estimated in Sect. \ref{Sect:LCshape}
		\\[1.5ex]
	\end{footnotesize}
\end{table*}

\begin{table*}
	\caption{Absolute magnitudes for 40 SNe in the $r$ band (corrected for distance modulus and MW extinction), compared with the absolute magnitudes in the $r$ band obtained when including the contribution from the host extinction with both methods, from the Na I D absorption and the $g-r$ method.}\label{Table:Mag_r}
	\begin{footnotesize}
		\begin{tabular}{lccc}
			\hline
			SN &  $M_{r}^{\rm{peak}}$  &   $M_{r}^{\rm{peak}}$ & $M_{r}^{\rm{peak}}$\\
			   & no EBV$_{\rm{host}}$ & EBV$_{\rm{host}}$ from Na I D & EBV$_{\rm{host}}$ from $g-r$\\
			& (mag) & (mag)  & (mag)\\
			\hline
		  09dh &  -18.88 & -19.30  & -19.25  \\
  09ut &  -16.53 & -16.53  & -17.16  \\
  10bip &  -17.52 & -17.67 &  -17.52 \\
  10hfe &  -16.75 & -16.75 &  -16.75 \\
  10hie &  -17.84 & -17.89 &  -18.68 \\
  10lbo &  -17.56 & -17.56 &  -17.56 \\
  10osn &  -16.90 & -16.90 &  -17.05 \\
  10tqi &  -17.15 & -17.57 &  -17.62 \\
  10yow &  -16.78 & -17.23 &  -17.33 \\
  10zcn &  -16.55 & -16.85 &  -18.19 \\
  11bli &  -17.42 & -17.67 &  -18.69 \\
  11bov &  -18.32 & -18.44 &  -18.32 \\
  11jgj &  -17.32 & -17.90 &  -18.00 \\
  11klg &  -17.26 & -17.26 &  -17.99 \\
  11mnb &  -18.29 & -18.49 &  -18.50 \\
  11mwk &  -18.39 & -18.44 &  -18.39 \\
  11rka &  -18.64 & -18.85 &  -19.40 \\
  12cjy &  -16.93 & -16.93 &  -16.93 \\
  12dcp &  -17.09 & -17.09 &  -18.07 \\
  12dtf &  -17.67 & -17.78 &  -18.56 \\
  12fgw &  -17.77 & -17.77 &  -17.77 \\
  12gty &  -19.73 & -19.81 &  -19.84 \\
  12gzk &  -18.65 & -19.01 &  -19.19 \\
  12hvv &  -17.37 & -17.57 &  -19.18 \\
  12jxd &  -16.04 & -16.87 &  -17.14 \\
  12ktu &  -16.38 & -16.59 &  -17.25 \\
  13ab &  -16.91 & -17.02  & -16.91  \\
  13aot &  -16.90 & -17.00 &   -- \\
  13cuv &  -17.26 & -17.30 &  -17.94 \\
  13dht &  -16.87 & -17.03 &  -17.38 \\
  13djf &  -15.45 & -15.53 &  -16.92 \\
  14bpy &  -17.78 & -17.78 &  -17.78 \\
  14fuz &  -17.40 & -17.52 &  -17.41 \\
  14gqr &  -17.41 & -17.41 &  -17.41 \\
  14jhf &  -17.82 & -18.62 &   -- \\
  14ym &  -17.37 & -17.76  & -17.94  \\
  15acp &  -18.77 & -19.10 &  -19.62 \\
  15dtg &  -18.00 & -18.00 &  -18.28 \\
  16flq &  -17.90 & -18.51 &  -18.49 \\
  16hgp &  -18.48 & -18.48 &  -18.65 \\	
				\hline
		\end{tabular}
	\\[1.5ex]
\end{footnotesize}
\end{table*}

\begin{table*}
	\caption{Estimated explosions time ($t_{explo}$), the first detection ($t_{firstdet}$), estimated rise times in the $r$ band ($t_{rise}(r)$) compared with the ones from the bolometric light curves ($t_{rise}(bolo)$) for the 44 SNe of our sample.}\label{Table:texplo_trise}
	\begin{footnotesize}
		\begin{tabular}{lcccc}
			\hline
			SN & $t_{explo}$ & $t_{firstdet}$ & $t_{rise}(r)$ & $t_{rise}(bolo)$  \\
			& (JD) & (JD)  & (days) & (days)\\
			\hline
  09dh & 2454929.02 & 2454938.80 & 23.75 & 20.70  \\
  09ut & 2454994.62 & 2455005.82 & 18.88 & 35.54  \\
  10bip & 2455213.75 & 2455221.85 & 17.09 & 13.90 \\ 
  10hfe & 2455324.83 & 2455331.87 & 15.57 & 12.44 \\ 
  10hie & 2455323.06 & 2455335.96 & 12.38 & 15.80 \\ 
  10lbo & 2455349.10 & 2455352.81 & 26.59 & 21.73 \\ 
  10osn & 2455385.78 & 2455388.98 & 11.95 & 10.89 \\ 
  10tqi & 2455426.96 & 2455429.98 & 16.00 & 13.66 \\ 
  10yow & 2455467.01 & 2455467.61 & 13.41 & 12.36 \\ 
  10zcn & 2455471.76 & 2455478.82 & 15.54 & 16.87 \\ 
  11bli & 2455629.49 & 2455635.75 & 15.97 & 17.13 \\ 
  11bov & 2455648.11 & 2455651.71 & 36.34 & 25.13 \\ 
  11hyg & 2455735.62 & 2455738.92 & 14.82 & 14.96 \\ 
  11jgj & 2455764.66 & 2455765.76 & 20.85 & 17.90 \\ 
  11klg & 2455791.67 & 2455792.87 & 16.94 & 15.47 \\ 
  11lmn & 2455794.29 & 2455801.74 & 16.97 & 21.19 \\ 
  11mnb & 2455804.34 & 2455809.88 & 53.77 & 54.87 \\ 
  11mwk & 2455811.23 & 2455820.83 & 16.55 & 14.31 \\ 
  11rka & 2455895.61 & 2455896.01 & 32.54 & 33.67 \\ 
  12cjy & 2455999.74 & 2456017.94 & 25.28 & 21.74 \\ 
  12dcp & 2456027.93 & 2456033.96 & 19.62 & 28.69 \\ 
  12dtf & 2456033.35 & 2456036.95 & 21.72 & 24.95 \\ 
  12fgw & 2456077.64 & 2456077.74 & 14.90 & 13.62 \\ 
  12gty & 2456086.82 & 2456086.82 & 72.38 & 52.64 \\ 
  12gzk & 2456132.86 & 2456137.82 & 19.31 & 16.55 \\ 
  12hvv & 2456148.97 & 2456151.87 & 16.51 & 28.63 \\ 
  12jxd & 2456206.04 & 2456226.04 & 23.31 & 21.50 \\ 
  12ktu & 2456227.59 & 2456231.89 & 17.24 & 19.50 \\ 
  13ab & 2456308.78 & 2456327.88 & 23.89 & 19.34  \\
  13aot & 2456394.05 & 2456395.85 & 22.09 & -- \\ 
  13cuv & 2456510.40 & 2456511.90 & 25.75 & 26.98 \\ 
  13dht & 2456531.98 & 2456542.68 & 18.64 & 16.41 \\ 
  13djf & 2456542.76 & 2456543.76 & 14.80 & 12.09 \\ 
  14bpy & 2456821.20 & 2456823.70 & 13.78 & 11.19 \\ 
  14fuz & 2456912.89 & 2456925.88 & 13.88 & 16.83 \\ 
  14gao & 2456911.17 & 2456912.97 & 14.85 & 13.73 \\ 
  14gqr & 2456944.25 & 2456944.34 & 7.18 & 6.06   \\
  14jhf & 2457008.50 & 2457011.90 & 25.08 & -- \\ 
  14ym & 2456692.23 & 2456711.03 & 22.48 & 18.23  \\
  15acp & 2457105.67 & 2457107.97 & 14.56 & 12.74 \\ 
  15cpq & 2457268.87 & 2457270.87 & 15.69 & 12.95 \\ 
  15dtg & 2457333.45 & 2457333.93 & 37.41 & 19.88 \\ 
  16flq & 2457617.46 & 2457627.96 & 25.57 & 19.34 \\ 
  16hgp & 2457678.69 & 2457681.79 & 30.90 & 21.76 \\
			\hline
		\end{tabular}
		\\[1.5ex]
	\end{footnotesize}
\end{table*}

\begin{table*}
	\caption{Estimated \ion{Fe}{II}~$\lambda$5169 velocities at peak for 44 SNe of the sample}\label{Table:vel_max}
	\begin{footnotesize}
		\begin{tabular}{lc}
			\hline
			SN & $v_{max}$ \\
			& ($\mathrm{km} \: \mathrm{s}^{-1}$) \\
			\hline
  09dh & 8125.6     \\
  09ut & 8125.6     \\
  10bip  & 13706.70 \\
  10hfe  & 9497.55  \\
  10hie  & 10108.39 \\
  10lbo  & 9140.21  \\
  10osn  & 4589.67  \\
  10tqi  & 9315.55  \\
  10yow  & 9464.53  \\
  10zcn  & 8819.06  \\
  11bli  & 4847.10  \\
  11bov  & 5815.18  \\
  11hyg  & 7832.18  \\
  11klg  & 8172.07  \\
  11jgj & 8125.6    \\
  11lmn  & 2437.49  \\
  11mnb  & 5005.96  \\
  11mwk  & 8573.56  \\
  11rka  & 5739.87  \\
  12cjy & 8125.6    \\
  12dcp  & 6493.21  \\
  12dtf  & 8894.58  \\
  12fgw & 8125.6    \\
  12gty  & 3492.84  \\
  12gzk  & 16984.08 \\
  12hvv  & 8076.82  \\
  12jxd  & 7831.30  \\
  12ktu  & 8319.47  \\
  13ab  & 11149.58  \\
  13aot  & 7728.21  \\
  13cuv  & 5389.32  \\
  13dht  & 7065.56  \\
  14djf & 8125.6    \\
  14bpy  & 11148.79 \\
  14fuz  & 9754.77  \\
  14gao & 10171.24  \\
  14gqr & 10472.35  \\
  14jhf  & 7989.49  \\
  14ym & 7436.66    \\
  15acp  & 6570.15  \\
  15cpq & 9960.42   \\
  15dtg  & 8261.14  \\
  16flq  & 6281.47  \\
  16hgp  & 6235.76  \\
				\hline
		\end{tabular}
		\\[1.5ex]
	\end{footnotesize}
\end{table*}

\begin{table*}
	\caption{Estimated explosion parameters for the 42 SNe. PTF12gty estimated explosion parameters are presented here in the table for completeness but this SN is excluded in the analysis.}  \label{Table:paramexplo}
	\begin{footnotesize}
		\begin{tabular}{lccc}
			\hline
			SN & $M_{ej}$ & $E_{K}$ & $M_{^{56}\mathrm{Ni}}$ \\
			& ($\msun$) & ($10^{51} \: \mathrm{erg}$) & ($\msun$)  \\
			\hline
  09dh &  4.83  (1.92) & 1.90 (0.38) & 0.56 (0.06)    \\
  09ut &  8.65  (3.44) & 3.41 (0.68) & 0.07 (0.01)    \\
  10bip &  3.99 (1.59) & 4.47 (0.89) & 0.13 (0.01)   \\
  10hfe &  1.23 (0.49) & 0.66 (0.13) & 0.09 (0.01)   \\
  10hie &  3.19 (1.27) & 1.95 (0.39) & 0.12 (0.01)   \\
  10lbo &  7.59 (3.02) & 3.78 (0.76) & 0.18 (0.02)   \\
  10osn &  0.58 (0.23) & 0.07 (0.01) & 0.05 (0.01)   \\
  10tqi &  2.49 (0.99) & 1.29 (0.26) & 0.09 (0.01)   \\
  10yow &  1.61 (0.64) & 0.86 (0.17) & 0.06 (0.01)   \\
  10zcn &  5.31 (2.11) & 2.47 (0.49) & 0.07 (0.01)   \\
  11bli &  1.82 (0.72) & 0.26 (0.05) & 0.16 (0.02)   \\
  11bov &  5.78 (2.30) & 1.17 (0.23) & 0.46 (0.05)   \\
  11hyg &  2.10 (0.83) & 0.77 (0.15) & 0.11 (0.01)   \\
  11jgj &  3.97 (1.58) & 1.56 (0.31) & 0.15 (0.01)   \\
  11klg &  2.54 (1.01) & 1.01 (0.20) & 0.08 (0.01)   \\
  11lmn &  0.56 (0.22) & 0.02 (0.01) & 0.15 (0.01)   \\
  11mnb &  31.29 (12.45) & 4.68 (0.94) & 0.74 (0.08) \\
  11mwk &  2.77 (1.10) & 1.22 (0.24) & 0.28 (0.03)   \\
  11rka &  11.05 (4.40) & 2.17 (0.43) & 0.63 (0.06)  \\
  12cjy &  3.77 (1.50) & 1.49 (0.30) & 0.08 (0.01)  \\
  12dcp &  4.15 (1.65) & 1.05 (0.21) & 0.10 (0.01)  \\
  12dtf &  7.04 (2.80) & 3.32 (0.66) & 0.19 (0.02)  \\
  12fgw &  1.18 (0.47) & 0.46 (0.09) & 0.24 (0.02)  \\
  12gzk &  6.31 (2.51) & 10.86( 2.17) & 0.41 ( 0.04)  \\
  12gty$^{a}$ &  21.42 (8.52) & 1.56 (0.31) & 2.98 (0.30) \\ 
  12hvv &  10.30 (4.10) & 4.01 (0.80) & 0.22 (0.02)  \\
  12jxd &  0.42 (0.17) & 0.15 (0.03) & 0.06 (0.01)   \\
  12ktu &  4.02 (1.60) & 1.66 (0.33) & 0.04 (0.02)   \\
  13ab &  1.68 (0.67) & 1.25 (0.25) & 0.11 (0.01)    \\
  13cuv &  5.19 (2.06) & 0.90 (0.18) & 0.12 (0.01)   \\
  13dht &  3.43 (1.36) & 1.02 (0.20) & 0.07 (0.01)   \\
  13djf &  3.78 (1.50) & 1.49 (0.30) & 0.02 (0.01)   \\
  14bpy &  1.63 (0.65) & 1.21 (0.24) & 0.13 (0.01)   \\
  14fuz &  3.76 (1.49) & 2.13 (0.43) & 0.10 (0.01)   \\
  14gao &  1.42 (0.56) & 0.87 (0.17) & 0.06 (0.01)   \\
  14gqr &  0.12 (0.05) & 0.08 (0.02) & 0.06 (0.01)   \\
  14ym &  5.79 (2.30) & 1.91 (0.38) & 0.15 (0.02)    \\
  15acp &  1.62 (0.64) & 0.42 (0.08) & 0.47 (0.05)   \\
  15cpq &  1.43 (0.57) & 0.85 (0.17) & 0.28 (0.03)   \\
  15dtg &  5.69 (2.26) & 2.32 (0.46) & 0.24 (0.02)   \\
  16flq &  7.64 (3.04) & 1.80 (0.36) & 0.36 (0.04)   \\
  16hgp &  3.01 (1.20) & 0.70 (0.14) & 0.33 (0.03)   \\
			\hline
		\end{tabular} \\
		$^{a}$ PTF12gty estimated explosion parameters are presented here in the table for completeness but this SN is excluded in the analysis.
		\\[1.5ex]
	\end{footnotesize}
\end{table*}

\begin{table*}
	\caption{Comparison of the average estimates of the explosions parameters with estimates from the literature.\label{Table:paramexplocomp}}
	\begin{footnotesize}
		\begin{tabular}{lccc}
			\hline
			 & $M_{ej}$ & $E_{K}$ & $M_{^{56}\mathrm{Ni}}$ \\
			& ($\msun$) & ($10^{51} \: \mathrm{erg}$) & ($\msun$)  \\
			\hline
Drout+11    & $1.7^{+1.4}_{-0.9}$ & $1.0^{+0.9}_{-0.5}$  & 0.24 (0.15)             \\
Taddia+15   & 5.7  (3.6)          & 1.7 (0.4)            & 0.33 (0.11)             \\
Lyman+16    & 3.0 (2.8)           & 1.9 (1.3)            & 0.22 (0.16)             \\
Prentice+16 & ...                 & ...                  & $0.16^{+0.03}_{-0.10}$  \\
Taddia+18   & 2.1 (1.0)           & 1.2 (0.7)            & 0.13 (0.04) \\
Prentice+19 & 3.0 (0.7)  & ...                  & 0.11 (0.09) \\
This work$^a$ & 4.50 (0.79) & 1.79 (0.29) & 0.19 (0.03) \\
This work$^b$ & 3.57 (0.40) & 1.74 (0.33) & 0.16 (0.02) \\
This work$^c$ & 3.67 (0.39) & 1.78 (0.32) & 0.16 (0.02) \\
This work$^d$ & 4.65 (0.92) & 1.88 (0.34) & 0.19 (0.03) \\
		\hline 
		\end{tabular} \\
		$^a$ Avarage values for the overall sample of 41 SNe \\
		$^b$ Average values when the 5 SNe with broad light curve are excluded \\
		$^c$ Average values when the 5 SNe with broad light curve and the 1 fast SN are excluded \\
	    $^d$ Average values when the 7 SNe with Ibc classification are excluded
		\\[1.5ex]
	\end{footnotesize}
\end{table*}

\end{document}